\documentclass[aps,prd,superscriptaddress,showpacs,preprintnumbers,amsmath,amssymb]{revtex4}

\usepackage{graphicx} 
\usepackage{dcolumn}  

\graphicspath{{ps}}

\def\EPJ{{Eur. Phys. Jour.} C}

\def\NPB{{Nucl. Phys. B\,}}
\def\PLB{{Phys. Lett.} B}

\def\PRD{{Phys. Rev.} D}
\def\PRL{{Phys. Rev. Lett.}}

\def\etal{{\it et al.}}

\usepackage{relsize}
\def\babar{\mbox{\slshape B\kern-0.1em{\smaller A}\kern-0.1em
    B\kern-0.1em{\smaller A\kern-0.2em R}}}
\def\pep2{PEP-II}

\mathchardef\Upsilon="7107
\def\Y#1S{\ensuremath{\Upsilon{(#1S)}}\xspace}

\RequirePackage{xspace}
\def\ra {\ensuremath{\rightarrow}\xspace}
\def\epem {\ensuremath{e^+e^-}\xspace}

\def\BB {\ensuremath{B\Bbar}\xspace}
\def\qqbar {\ensuremath{q\overline q}\xspace}

\def\Kpi{\mbox{$\Kmi\pipl$}}
\def\Ktwopi{\mbox{$\Kmi\pipl\pizero$}}
\def\Ktwopipl{\mbox{$\Kmi\pipl\pipl$}}
\def\Kthreepi{\mbox{$\Kmi\pipl\pipl\pimi$}}

\def\De{\mbox{$\Delta E$}}
\def\Mb{\mbox{$M_{\rm bc}$}}
\def\Eb{\mbox{$E_{\rm beam}^{*}$}}

\providecommand{\cm}{\,\mbox{\rm cm}}

\providecommand{\MeV}{\,\mbox{\rm MeV}}
\providecommand{\MeVc}{\,\mbox{${\rm MeV}/c$}}
\providecommand{\MeVcsq}{\,\mbox{${\rm MeV}/c^2$}}
\providecommand{\GeV}{\,\mbox{\rm GeV}}
\providecommand{\GeVc}{\,\mbox{${\rm GeV}/c$}}
\providecommand{\GeVcsq}{\,\mbox{${\rm GeV}/c^2$}}

\providecommand{\hzero}{\mbox{$h^0$}}
\providecommand{\hmi}{\mbox{$h^-$}}

\providecommand{\Wmi}{\mbox{$W^-$}}
\providecommand{\gamgam}{\gamma \gamma}
\providecommand{\pipl}{\mbox{$\pi^+$}}
\providecommand{\pimi}{\mbox{$\pi^-$}}
\providecommand{\pizero}{\mbox{$\pi^0$}}

\providecommand{\threepi}{\mbox{\pipl \pimi \pizero}}

\providecommand{\etap}{\mbox{$\eta^\prime$}}

\providecommand{\rhomi}{\mbox{$\rho^-$}}

\providecommand{\qbar}{\mbox{$\overline{q}$}}

\providecommand{\cbar}{\mbox{$\overline{c}$}}

\providecommand{\bbar}{\mbox{$\overline{b}$}}

\providecommand{\Kbar}{\kern 0.18em\overline{\kern -0.18em
K}{}\xspace}

\providecommand{\Kmi}{\mbox{$K^-$}}

\providecommand{\Kbar}{\mbox{$\overline{K}$}}

\providecommand{\bbar}{\mbox{\bbar}}
\providecommand{\cbar}{\mbox{\cbar}}
\providecommand{\qbar}{\mbox{\qbar}}

\providecommand{\Dzero}{\mbox{$D^0$}}

\providecommand{\Dpl}{\mbox{$D^+$}}

\providecommand{\Dstarpl}{\mbox{$D^{*+}$}}

\providecommand{\Dstarzero}{\mbox{$D^{*0}$}}

\providecommand{\Dstpl}{\mbox{$D^{(*)+}$}}
\providecommand{\Dstze}{\mbox{$D^{(*)0}$}}
\providecommand{\B}{\mbox{$B$\,}}

\providecommand{\Bbar}{\kern 0.18em\overline{\kern -0.18em
B}{}\xspace}

\providecommand{\Bzero}{\mbox{$B^0$}}
\providecommand{\Bzerobar}{\mbox{$\Bbar^0$}}

\providecommand{\Bmi}{\mbox{$B^-$}}

\providecommand{\tabbls}{0.8}
\providecommand{\tabelemsize}{\footnotesize}

\def\mod#1{#1}

\usepackage{color}
\def\refmodcolor{black}
\def\refmod#1{\textcolor{\refmodcolor}{#1}}

\begin{document}

\title{ \quad\\[1.0cm] Improved Measurements of Color-Suppressed Decays \\
$\Bzerobar \to \Dzero \pizero $, $\Dzero \eta$, $\Dzero \omega$,  
$\Dstarzero \pizero$, $\Dstarzero \eta$ and  $\Dstarzero \omega$ 
}

\affiliation{Budker Institute of Nuclear Physics, Novosibirsk}
\affiliation{Chiba University, Chiba}
\affiliation{Chonnam National University, Kwangju}
\affiliation{University of Cincinnati, Cincinnati, Ohio 45221}
\affiliation{The Graduate University for Advanced Studies, Hayama, Japan} 
\affiliation{University of Hawaii, Honolulu, Hawaii 96822}
\affiliation{High Energy Accelerator Research Organization (KEK), Tsukuba}
\affiliation{University of Illinois at Urbana-Champaign, Urbana, Illinois 61801}
\affiliation{Institute of High Energy Physics, Chinese Academy of Sciences, Beijing}
\affiliation{Institute of High Energy Physics, Vienna}
\affiliation{Institute of High Energy Physics, Protvino}
\affiliation{Institute for Theoretical and Experimental Physics, Moscow}
\affiliation{J. Stefan Institute, Ljubljana}
\affiliation{Kanagawa University, Yokohama}
\affiliation{Korea University, Seoul}
\affiliation{Kyungpook National University, Taegu}
\affiliation{Swiss Federal Institute of Technology of Lausanne, EPFL, Lausanne}
\affiliation{University of Ljubljana, Ljubljana}
\affiliation{University of Maribor, Maribor}
\affiliation{University of Melbourne, Victoria}
\affiliation{Nagoya University, Nagoya}
\affiliation{Nara Women's University, Nara}
\affiliation{National Central University, Chung-li}
\affiliation{National United University, Miao Li}
\affiliation{Department of Physics, National Taiwan University, Taipei}
\affiliation{H. Niewodniczanski Institute of Nuclear Physics, Krakow}
\affiliation{Nippon Dental University, Niigata}
\affiliation{Niigata University, Niigata}
\affiliation{University of Nova Gorica, Nova Gorica}
\affiliation{Osaka City University, Osaka}
\affiliation{Osaka University, Osaka}
\affiliation{Panjab University, Chandigarh}
\affiliation{Peking University, Beijing}
\affiliation{Princeton University, Princeton, New Jersey 08544}
\affiliation{RIKEN BNL Research Center, Upton, New York 11973}
\affiliation{Saga University, Saga}
\affiliation{University of Science and Technology of China, Hefei}
\affiliation{Seoul National University, Seoul}
\affiliation{Sungkyunkwan University, Suwon}
\affiliation{University of Sydney, Sydney NSW}
\affiliation{Tata Institute of Fundamental Research, Bombay}
\affiliation{Toho University, Funabashi}
\affiliation{Tohoku Gakuin University, Tagajo}
\affiliation{Tohoku University, Sendai}
\affiliation{Department of Physics, University of Tokyo, Tokyo}
\affiliation{Tokyo Institute of Technology, Tokyo}
\affiliation{Tokyo Metropolitan University, Tokyo}
\affiliation{Tokyo University of Agriculture and Technology, Tokyo}
\affiliation{Virginia Polytechnic Institute and State University, Blacksburg, Virginia 24061}
\affiliation{Yonsei University, Seoul}
  \author{S.~Blyth}\affiliation{National Central University, Chung-li} 
  \author{K.~Abe}\affiliation{High Energy Accelerator Research Organization (KEK), Tsukuba} 
  \author{K.~Abe}\affiliation{Tohoku Gakuin University, Tagajo} 
  \author{I.~Adachi}\affiliation{High Energy Accelerator Research Organization (KEK), Tsukuba} 
  \author{H.~Aihara}\affiliation{Department of Physics, University of Tokyo, Tokyo} 
  \author{D.~Anipko}\affiliation{Budker Institute of Nuclear Physics, Novosibirsk} 
  \author{V.~Aulchenko}\affiliation{Budker Institute of Nuclear Physics, Novosibirsk} 
  \author{T.~Aushev}\affiliation{Institute for Theoretical and Experimental Physics, Moscow} 
  \author{A.~M.~Bakich}\affiliation{University of Sydney, Sydney NSW} 
  \author{V.~Balagura}\affiliation{Institute for Theoretical and Experimental Physics, Moscow} 
% ...
\collaboration{The Belle Collaboration}

\noaffiliation
\begin{abstract}
We present improved measurements of the branching fractions of the color-suppressed  
decays 
$\Bzerobar \to  \Dstze \hzero$, 
where $\hzero$ 
represents \mod{a light neutral meson $\pizero$, $\eta$ or $\omega$}.
The measurements are based on a data sample of 140 \ensuremath{ \mathrm{fb}^{-1}\,} collected at the 
\ensuremath{ \Upsilon(4S) }\, \mod{resonance} with the Belle detector at
the KEKB energy-asymmetric \ensuremath{e^+e^-}\, collider, corresponding to  
seven times the luminosity of the previous Belle measurements.  
All the measured branching fractions fall in the range 1.4-2.4 $\times 10^{-4}$,   
which is significantly higher than theoretical predictions based on naive factorization.
\end{abstract}


\maketitle

\tighten

{\renewcommand{\thefootnote}{\fnsymbol{footnote}}}
\setcounter{footnote}{0}

\section{Introduction}
\begin{figure}[htb]
\begin{center}
\begin{tabular}{cc}
\mod{(a)} & 
\mod{(b)} \\
\includegraphics[width=1.7in]{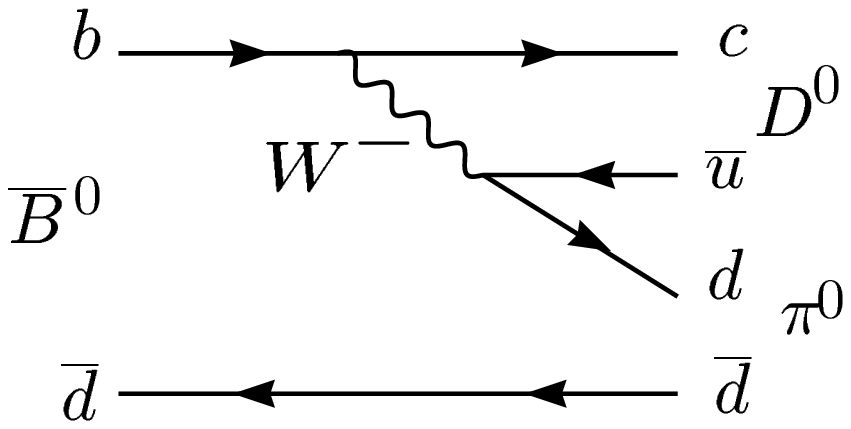}
 & 
\includegraphics[width=1.7in]{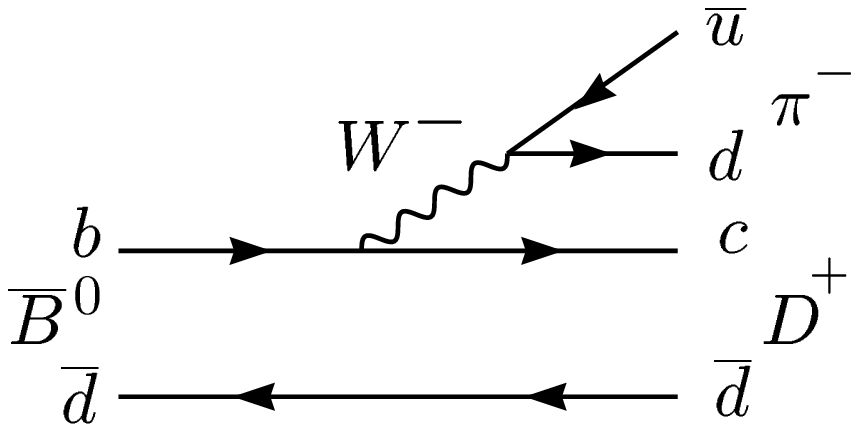}
 
\end{tabular}
\medskip
\caption{Tree level internal \mod{(a)} and external \mod{(b)} spectator 
diagrams for $\overline{B}\to D\pi$ decays.}
\label{dpi-feynman}
\end{center}
\end{figure}

The weak decays $\Bzerobar \ra \Dstze \hzero$~\cite{CC}, where $\hzero$ represents a light neutral 
meson, are expected to proceed predominantly through internal spectator diagrams, as illustrated
in Fig.~\ref{dpi-feynman}\mod{a}.
The color matching requirement between the quarks from the virtual $\Wmi$ and the
other quark pair results in these decays being ``color-suppressed'' relative to decays
such as $\Bzerobar \ra \Dstpl \hmi$, which proceed through external spectator
diagrams \mod{ as shown in Fig.~\ref{dpi-feynman}b }.

Previous measurements of 
$\Bzerobar$ decays into $\Dzero \rho^0$~\cite{ref:Belle2} and  
into $\Dstze \pizero$, $\Dzero \eta$ and $\Dzero \omega$ ~\cite{ref:Belle},
by the Belle collaboration, and of  
$\Bzerobar$ into $\Dstze \pizero$ by the CLEO collaboration~\cite{ref:CLEO}, 
and into $\Dstze \pizero$, $\Dstze \eta$ and $\Dstze \omega$
by the BaBar collaboration~\cite{ref:Babar} all indicate color-suppressed branching
fractions in the approximate range $(2$--$4)\times 10^{-4}$.
Further color-suppressed branching fraction measurements of  
$\Bzerobar$ decays into $\Dstze \etap$ by the Belle
collaboration~\cite{ref:BelleEtap} yield results of approximately $(1.1$--$1.2) \times 10^{-4}$.
Most of these measurements are substantially in excess of theoretical expectations from ``naive" factorization
models~[7--13]
\nocite{ref:Beneke,ref:NeuSte,ref:NeuPet,ref:Chua,ref:Rosner,ref:Deandrea,ref:ChRos},
which fall in the range \mod{ $(0.3$--$1.0) \times 10^{-4}$}. 

Several approaches to achieving a better theoretical description~\cite{ref:NeuPet,ref:Chua,ref:SCET,ref:pQCD} have been developed. 
They extend upon the factorization approach with consideration of 
final state interactions and consequent simultaneous treatment of  isospin amplitudes of  color-suppressed and color-allowed decays.
The possibility that similar effects could have dramatic implications
on 
direct $CP$ violation asymmetries in charmless decays, together with some degree of 
discrepancy between the prior Belle~\cite{ref:Belle} and BaBar~\cite{ref:Babar} measurements provide strong 
motivation for more precise measurements of the color-suppressed decays.

In this paper we report improved branching fraction measurements 
of  $\Bzerobar$ decays into 
$\Dzero \pizero$, 
$\Dzero \eta$, 
$\Dzero \omega$,  
$\Dstarzero \pizero$, 
$\Dstarzero \eta$ and 
$\Dstarzero \omega$. 
The measurements are  
based on a $140~{\rm fb}^{-1}$ data sample, which contains 152 million $B\overline{B}$ pairs, 
collected  with the Belle detector at the KEKB asymmetric-energy
$e^+e^-$ (3.5 on 8~GeV) collider~\cite{KEKB}
operating at the $\Upsilon(4S)$ resonance.  This corresponds to seven times the luminosity  
of the previous Belle measurements~\cite{ref:Belle} and almost twice that of the 
BaBar measurements~\cite{ref:Babar}. 
\refmod{Recent measurements of the same processes by the Belle Collaboration~\cite{ref:belledphi1} have been 
used to extract the angle $\phi_{1}$ of the CKM Unitarity Triangle 
using a time-dependent Dalitz analysis of $D\to K^{0}_{s} \pi^{+}\pi^{-}$,
allowing resolution of the sign ambiguities inherent in other determinations.}  

The Belle detector is a large-solid-angle magnetic
spectrometer that consists of a three-layer silicon vertex detector (SVD),
a 50-layer central drift chamber (CDC), an array of
aerogel threshold \v{C}erenkov counters (ACC), 
a barrel-like arrangement of time-of-flight
scintillation counters (TOF), and an electromagnetic calorimeter
comprised of CsI(Tl) crystals (ECL) located inside 
a superconducting solenoid coil that provides a 1.5~T
magnetic field.  An iron flux-return located outside of
the coil is instrumented to detect $K_L^0$ mesons and to identify
muons (KLM).  The detector
is described in detail elsewhere~\cite{Belle}.

\providecommand{\pifigcaption}{
Distributions of $\Mb$ and $\De$ for the decay modes $ \Bzerobar \to \Dzero\pizero$ (a,b) and $\Bzerobar \to \Dstarzero\pizero $ (c,d), 
with the three $\Dzero$ subdecay modes combined.  
The points represent the data, the
solid lines show the result of the fit and the dash-dotted lines  (peaking at $\De \approx 0 \GeV$ and $\Mb \approx 5.28 \GeVcsq$) 
represent the signal contributions. 
The vertical dotted lines represent the signal region.  
For the $\Mb$ distributions the upper \mod{long dashed line} shows the continuum-like background contribution, with peaking background contribution
represented by the \mod{lower long dashed line} and cross-feed contribution
represented by the bold dotted line.   
For the $\De$ distributions,  (b,d) the two \mod{long dashed curves} show the $B$
background components  \mod{(dominated by the color allowed $\Bmi \to \Dstze \rhomi  $ modes )} and the dotted line represents the continuum contribution   and the bold dotted curve represents the cross-feed contribution.
}

\begin{figure*}
\begin{tabular}{lr} 
\hspace{-1cm}
\includegraphics[width=0.35\textwidth]{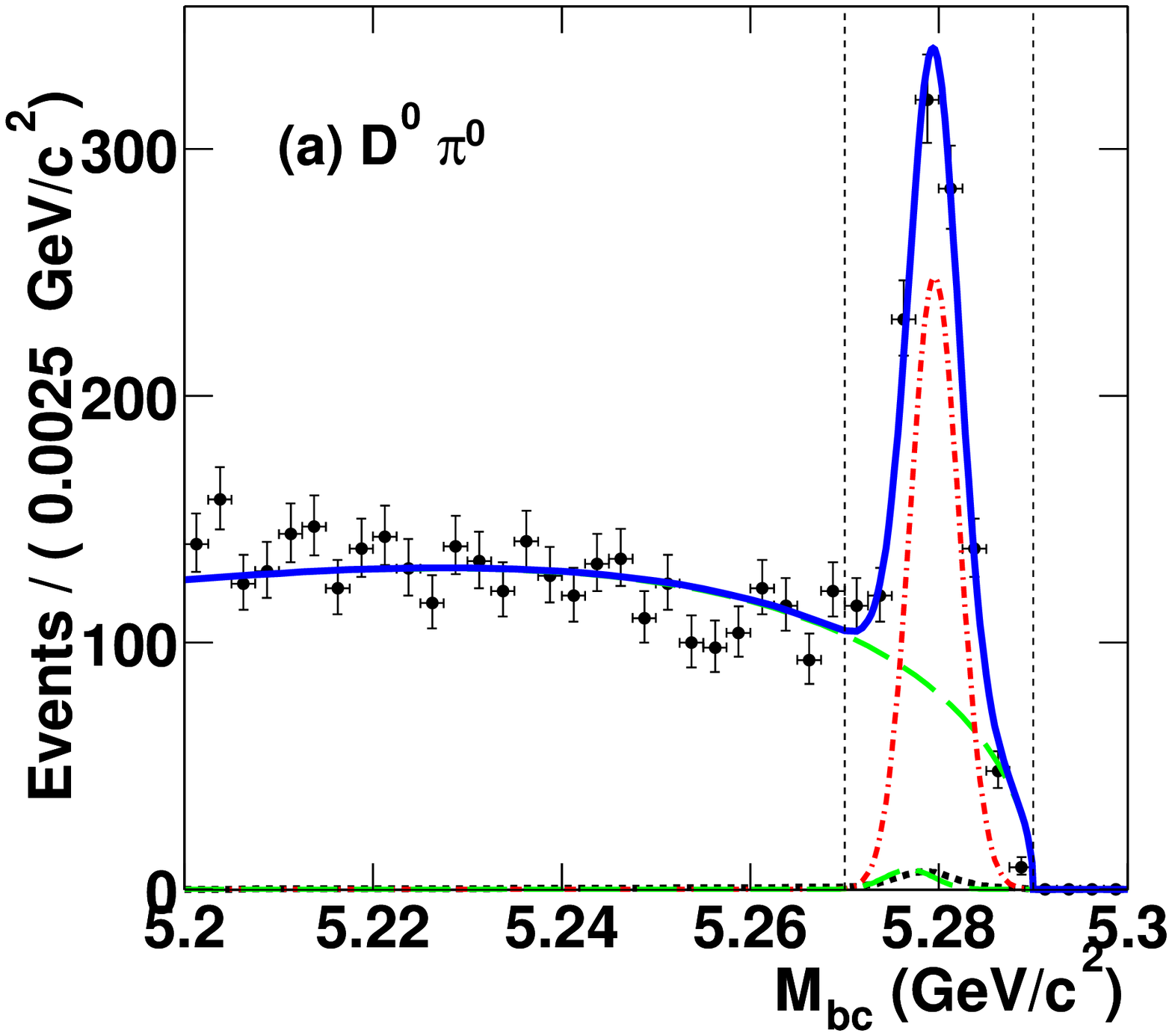}
 & 
\includegraphics[width=0.35\textwidth]{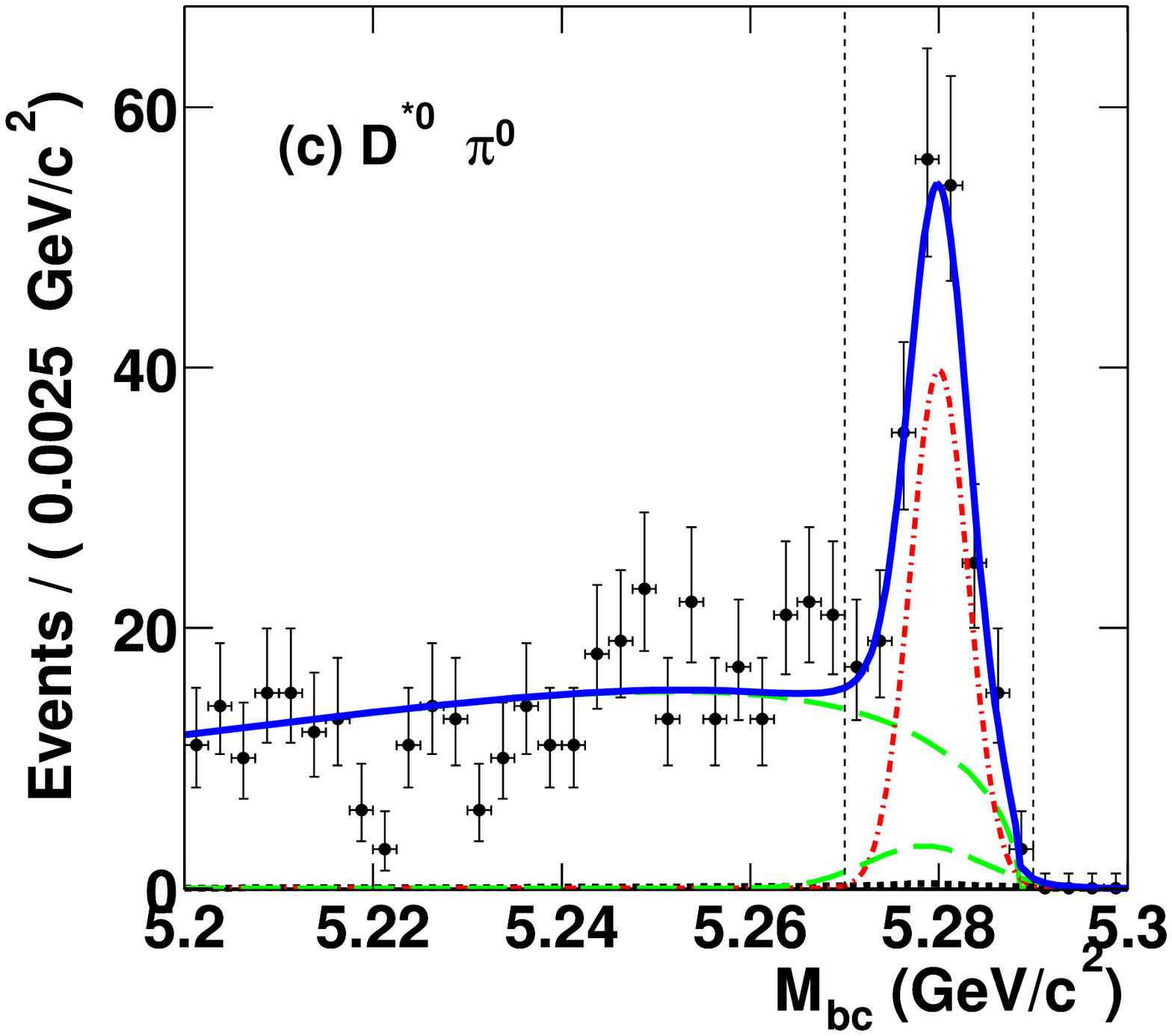}

\hspace{-0.25cm}
\\ 
\hspace{-1cm}
\includegraphics[width=0.35\textwidth]{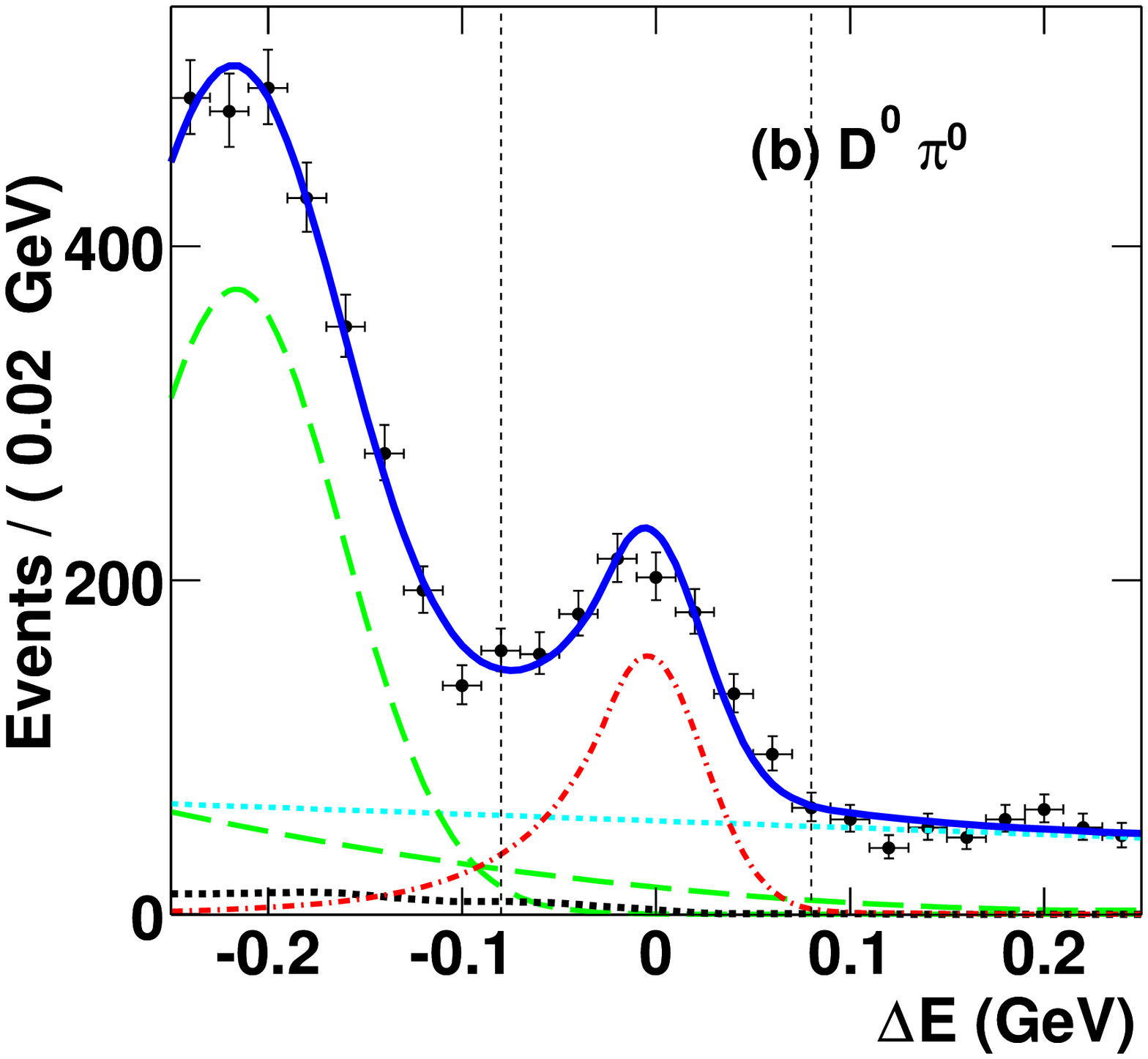}
 & 
\includegraphics[width=0.35\textwidth]{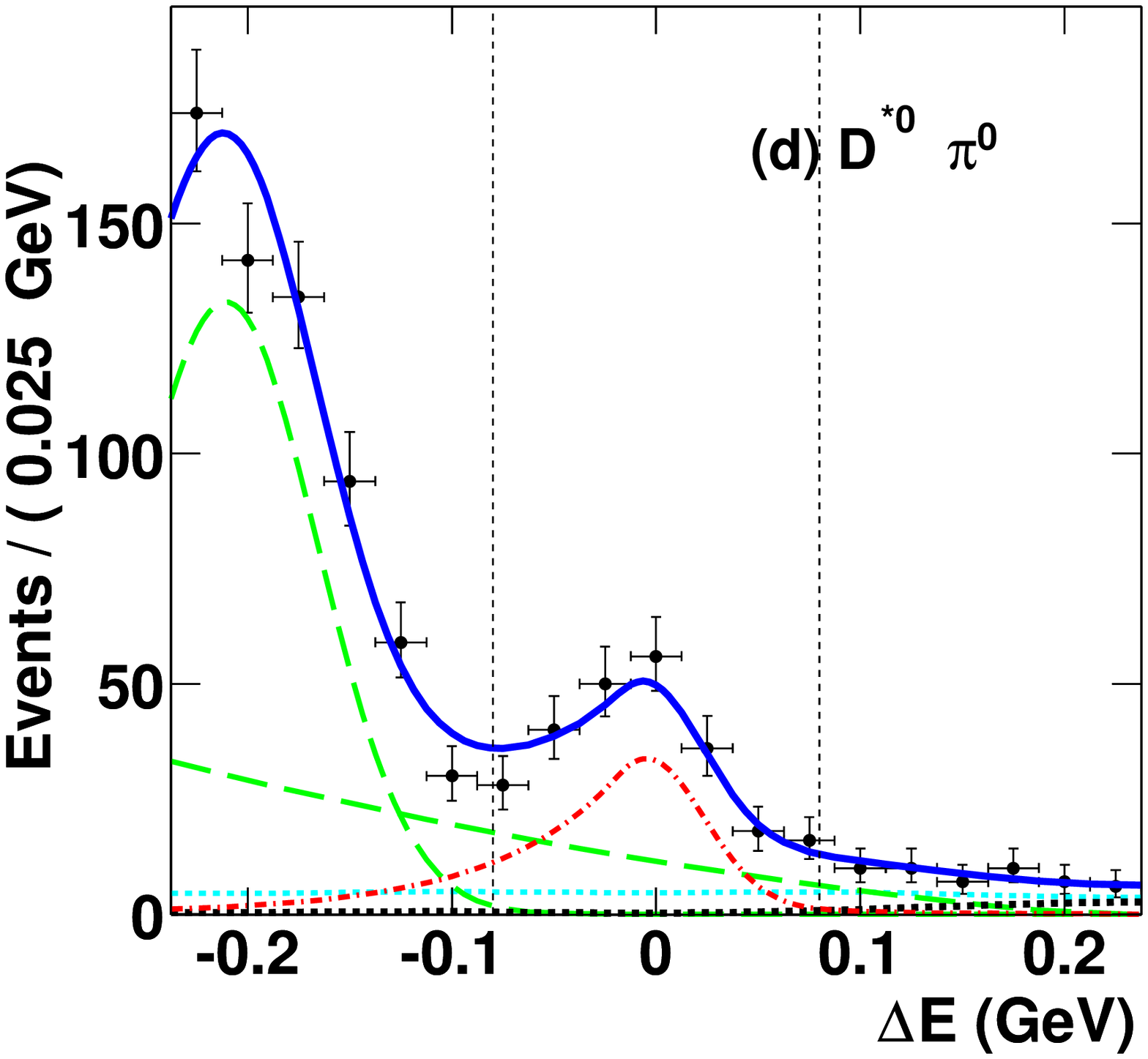}

\hspace{-0.25cm}
\end{tabular}
\begin{center}
\caption[]{
\pifigcaption
\label{fig:dstze_pi0}
}
\end{center}
\end{figure*}

\providecommand{\etafigcaption}{
Distributions of $\Mb$ and $\De$ for the decay modes $ \Bzerobar \to \Dzero\eta$ (a,b) and    $\Bzerobar \to \Dstarzero\eta $ (c,d),
with the  three $\Dzero$ subdecay modes combined. 
For the $\De$ distributions,   (b) the \mod{long dashed curve} represents the sum of $B$ background and continuum contributions; 
(d) the \mod{long dashed curve} represents the $B$ background contribution with the dotted line representing the continuum contribution;  
in both (b,d) the bold dotted curves represent the cross-feed contributions. 
\mod{In (b) two bold dotted curves are visible, corresponding to
cross-feed arising from $\Dstarzero\hzero$ final states;  
the broader distribution is the component   
with final states where  $\Dstarzero\to \Dzero \gamma$ and the other
corresponds to final states where  $\Dstarzero \to \Dzero \pizero$. 
} 
Other conventions follow those described in the caption of Figure~\ref{fig:dstze_pi0}.  
}

\begin{figure*}
\begin{tabular}{lr} 
\hspace{-1cm}
\includegraphics[width=0.35\textwidth]{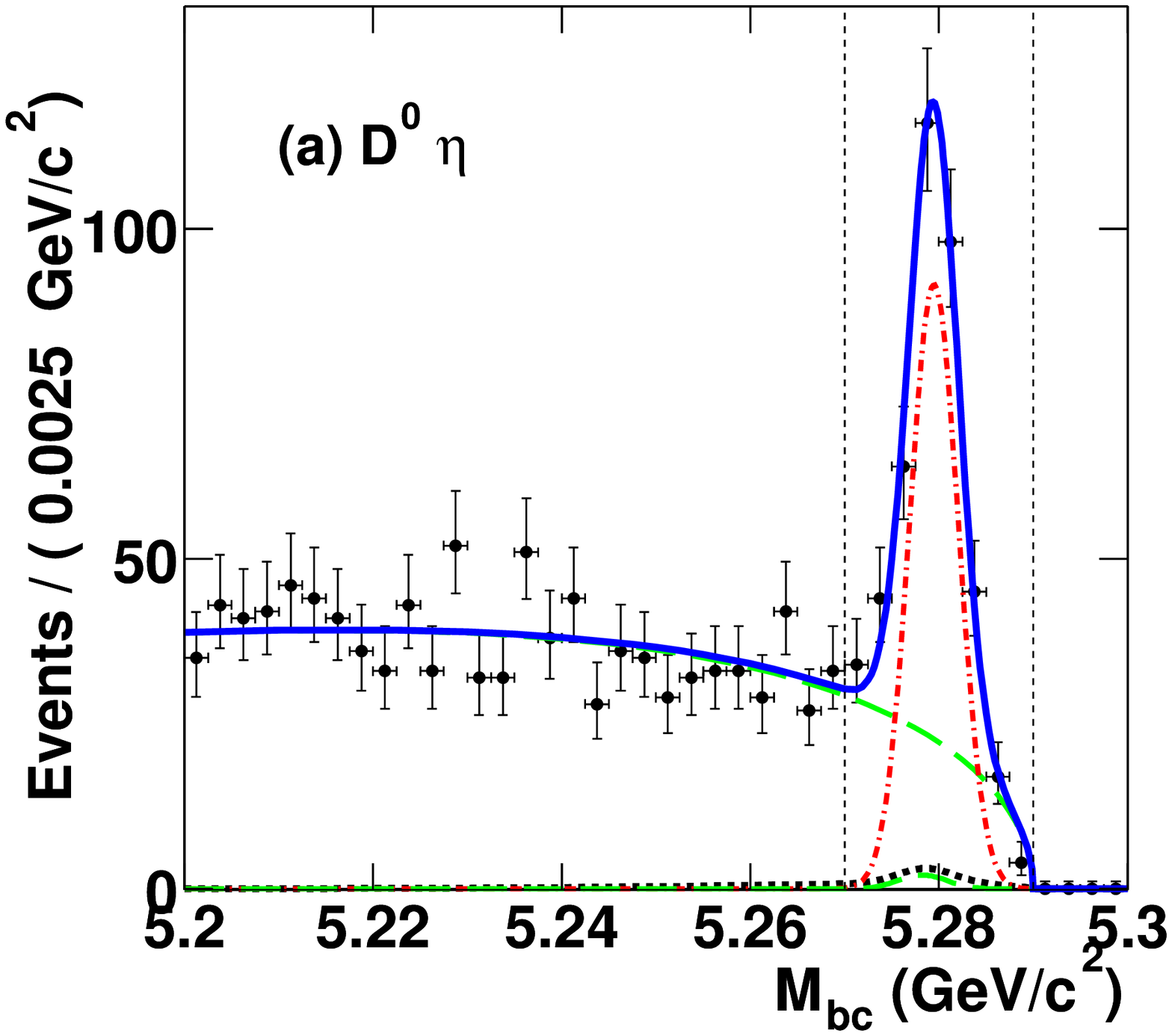}
 & 
\includegraphics[width=0.35\textwidth]{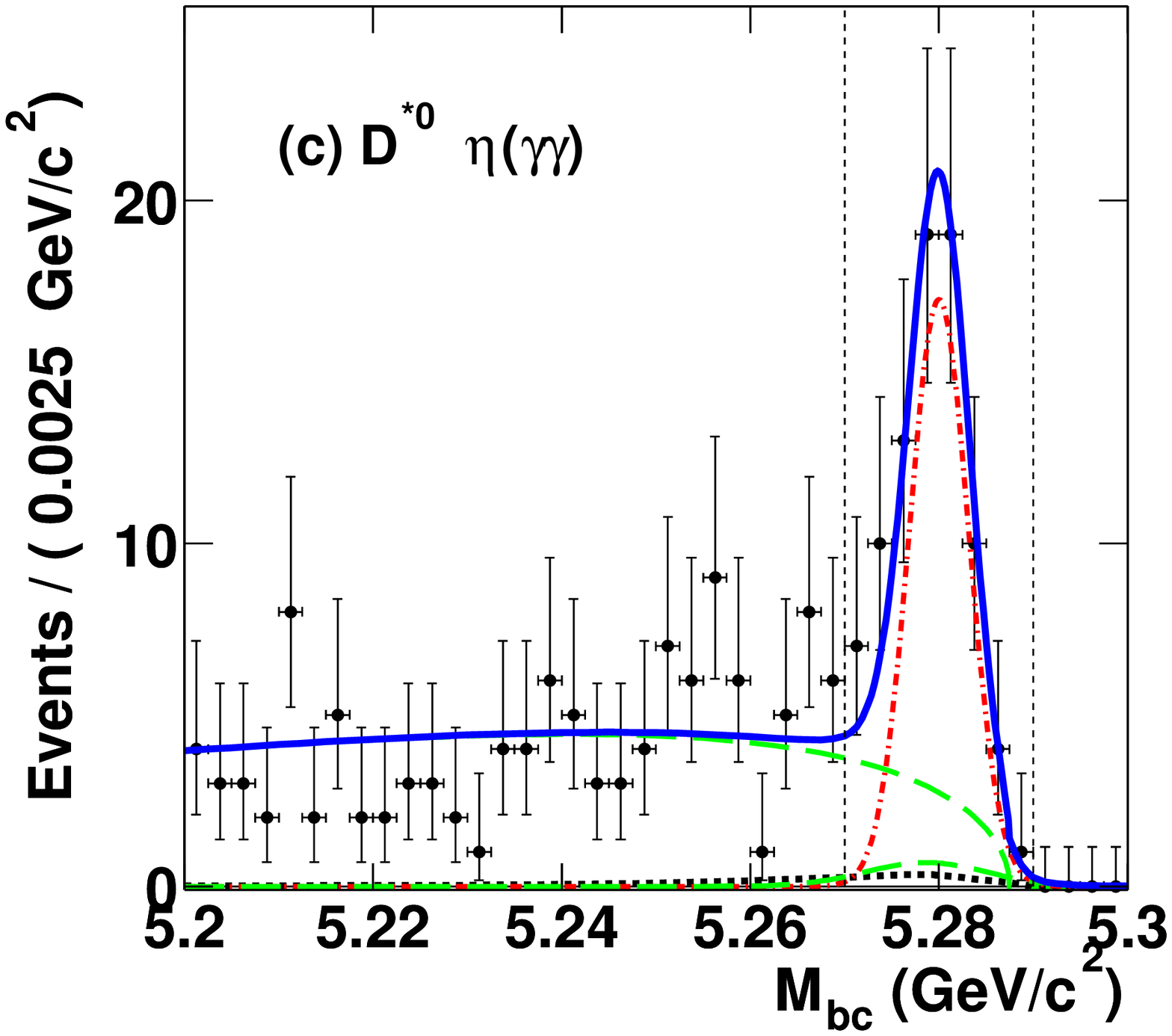}

\hspace{-0.25cm}
\\ 
\hspace{-1cm}
\includegraphics[width=0.35\textwidth]{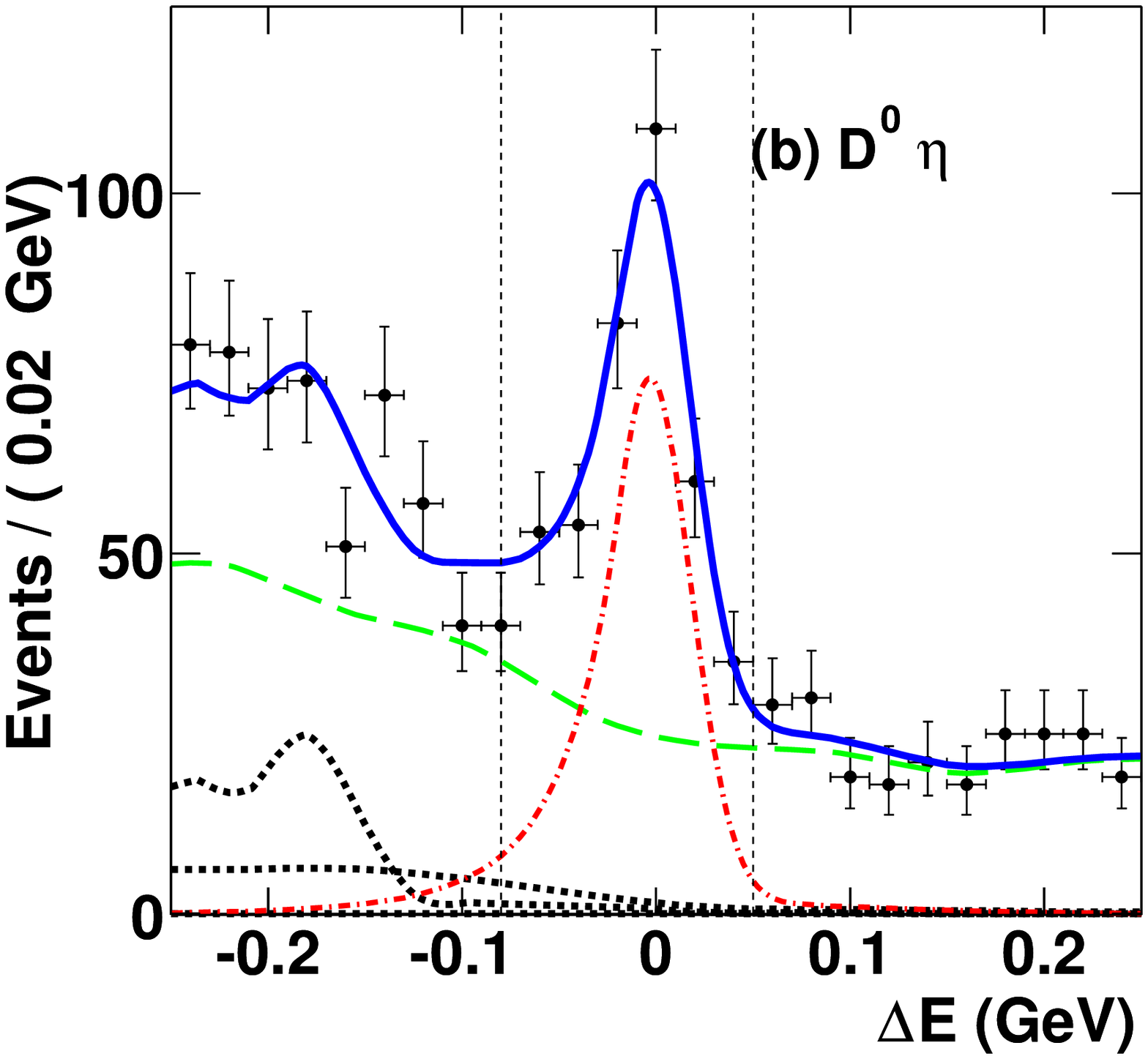}
 & 
\includegraphics[width=0.35\textwidth]{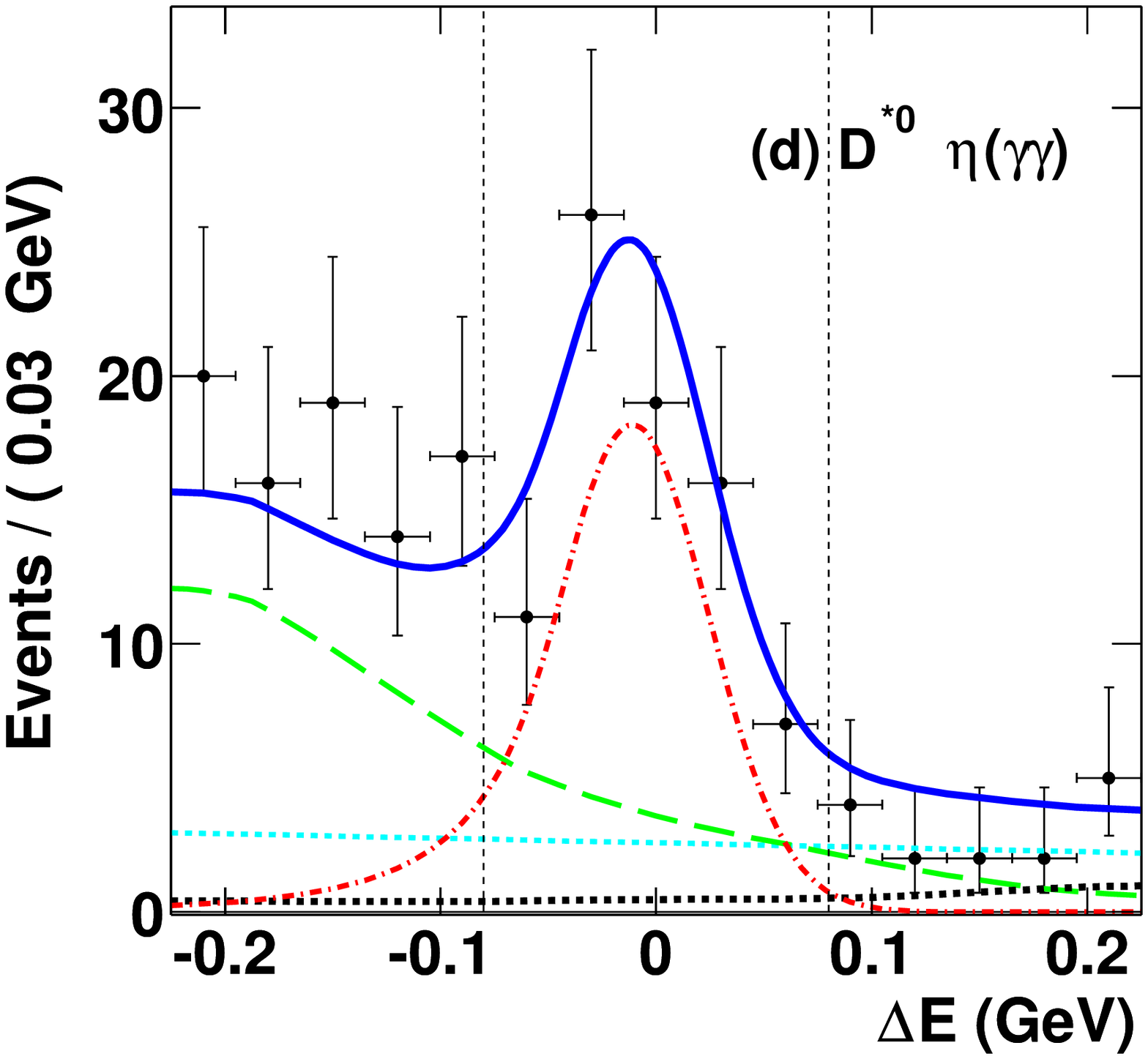}

\hspace{-0.25cm}
\end{tabular}
\begin{center}
\caption[]{
\etafigcaption
\label{fig:dstze_eta}
}
\end{center}
\end{figure*}

\providecommand{\omegafigcaption}{
Distributions of $\Mb$ and $\De$ for the decay modes $ \Bzerobar \to \Dzero\omega$ (a,b) and  $\Bzerobar \to \Dstarzero\omega $ (c,d) , with the
three $\Dzero$ subdecay modes combined. 
	\mod{
The conventions follow those of Figure~\ref{fig:dstze_eta}.
}
}

\providecommand{\cffigcaption}{
Comparison of the measured branching fractions ($\times 10^{-4}$) of the modes
$\Bzerobar \to \Dstze\hzero$,  for each of the three $\Dzero$ subdecay modes
and for the combined mode samples,  as obtained from the $\De$ fit, with the $\Mb$ signal region requirement applied. The shaded band indicates the combined submode result. The BaBar results~\cite{ref:Babar} from $\Mb$ fits are also shown.    
}

\begin{figure*}
\begin{tabular}{lr} 
\hspace{-1cm}
\includegraphics[width=0.35\textwidth]{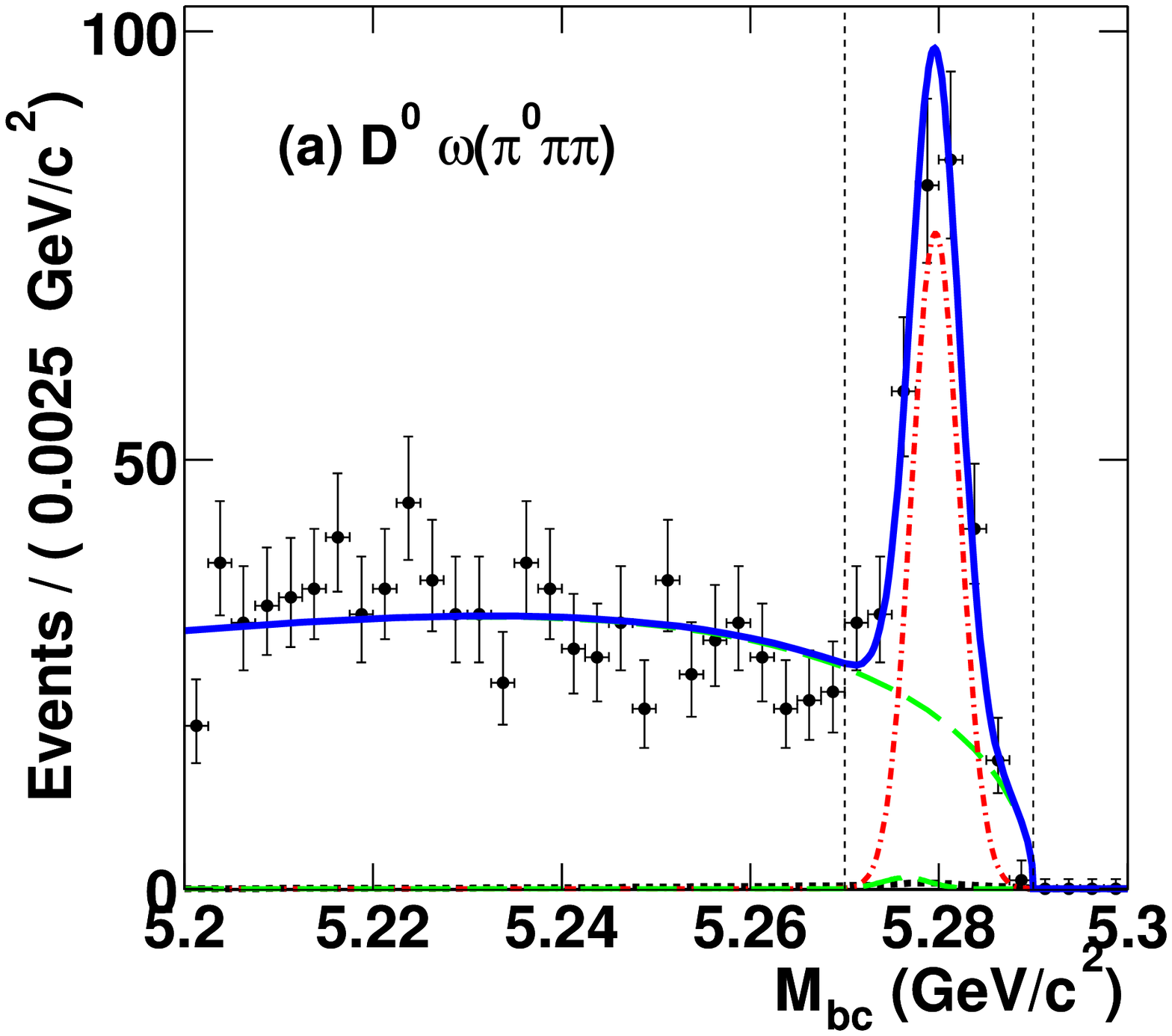}
 & 
\includegraphics[width=0.35\textwidth]{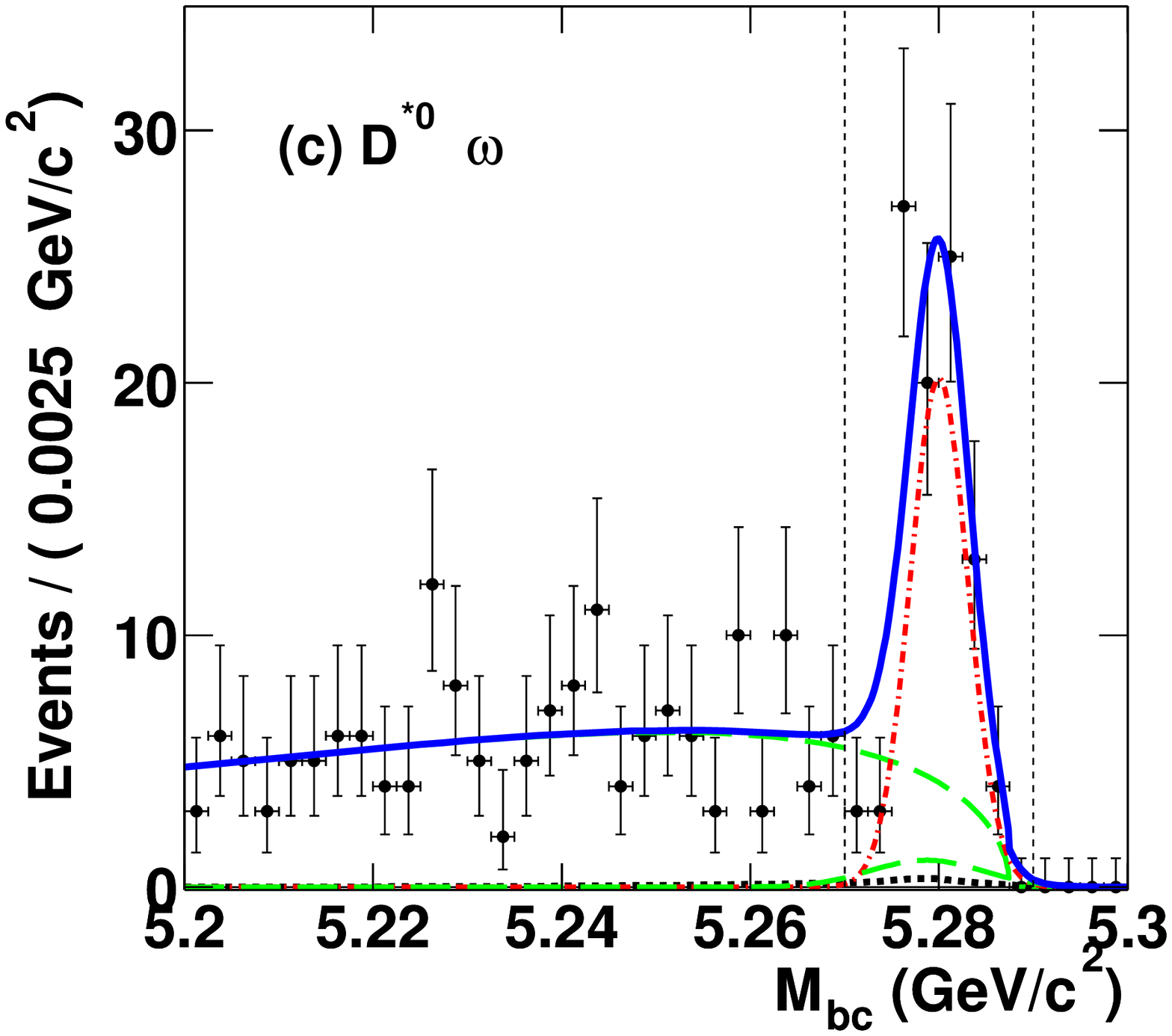}

\hspace{-0.25cm}
\\ 
\hspace{-1cm}
\includegraphics[width=0.35\textwidth]{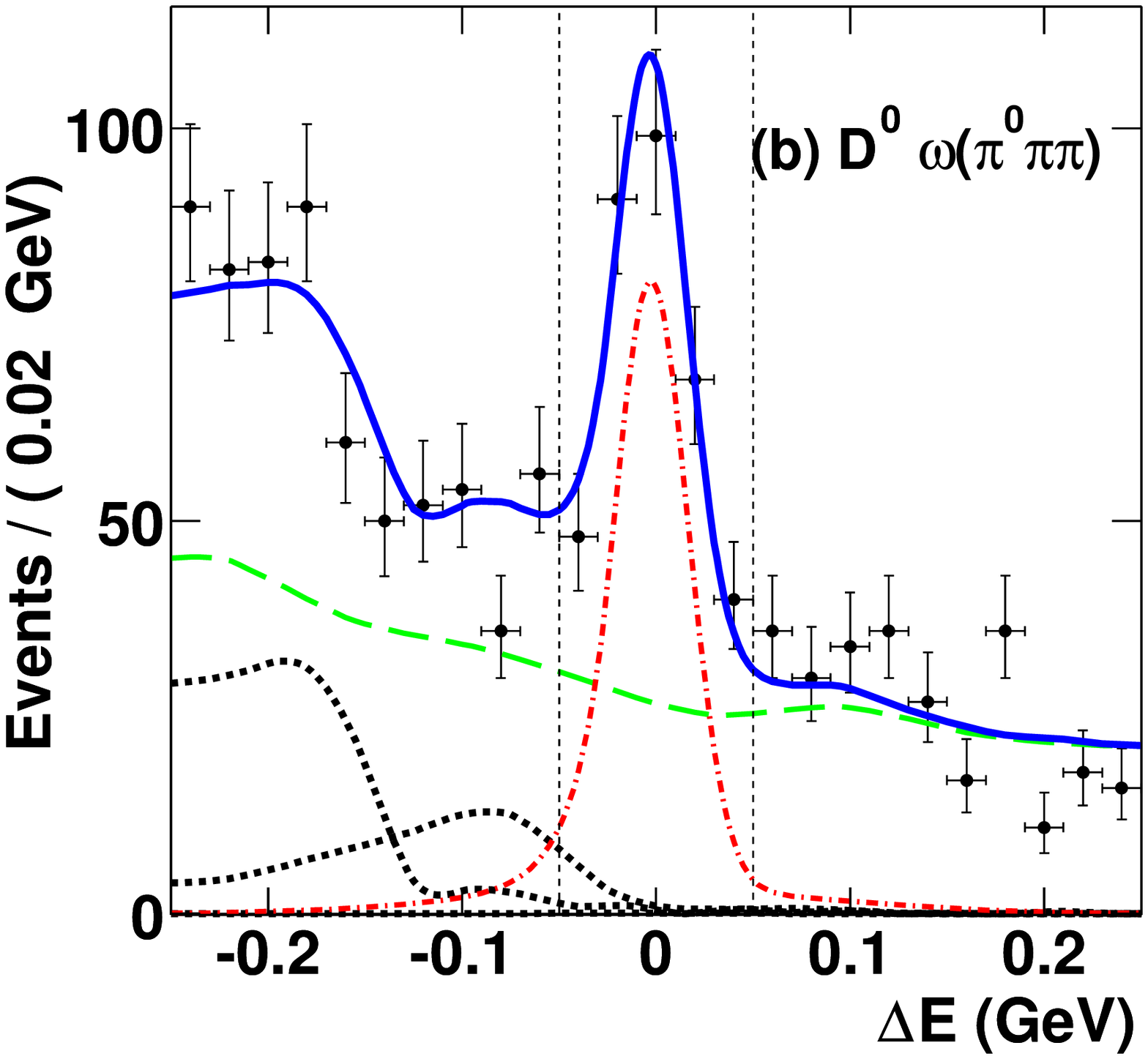}
 & 
\includegraphics[width=0.35\textwidth]{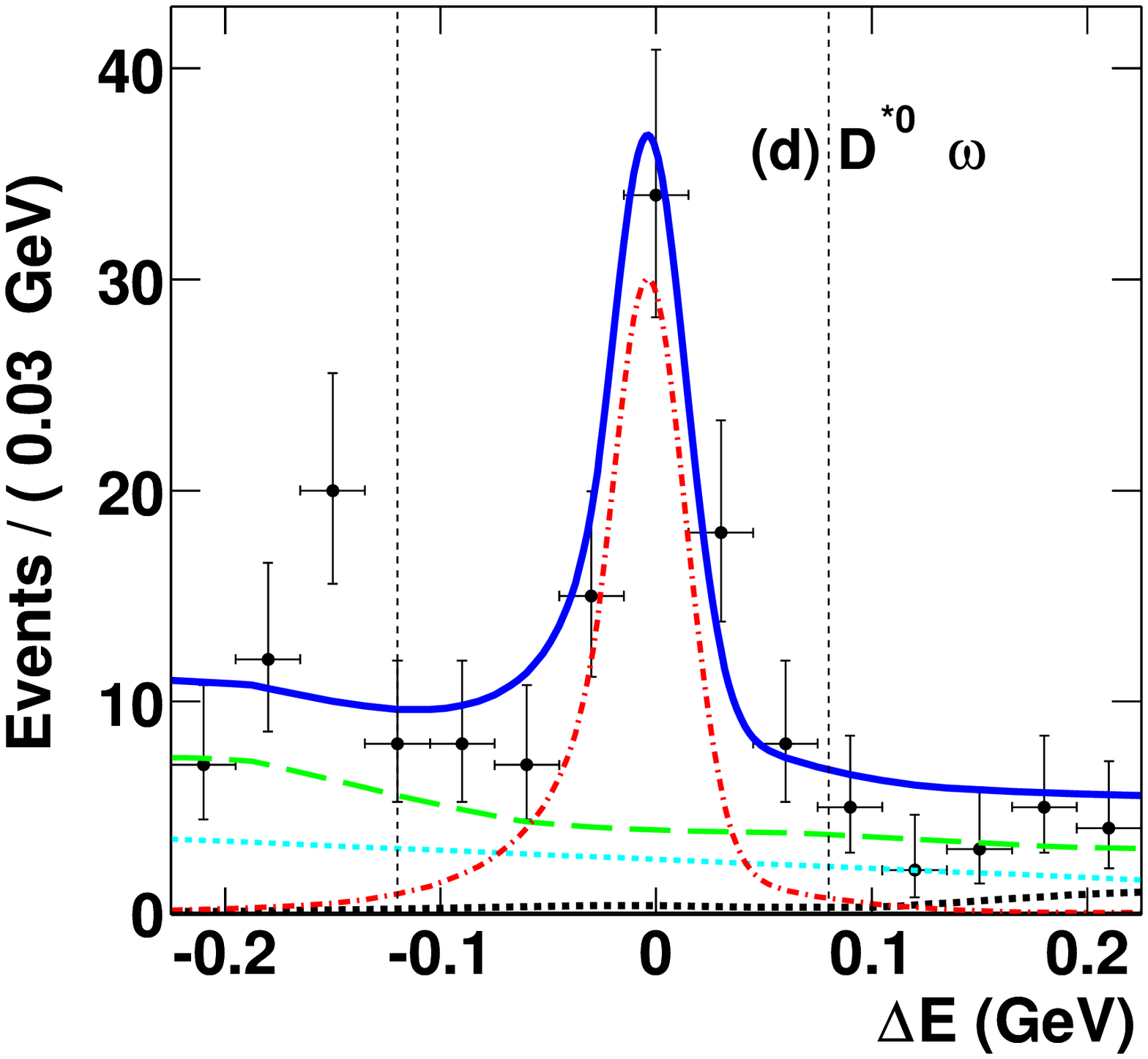}

\hspace{-0.25cm}
\end{tabular}
\begin{center}
\caption[]{
\omegafigcaption
\label{fig:dstze_omega}
}
\end{center}
\end{figure*}
\section{Event Selection}
Color-suppressed $\Bzerobar$ meson decays are reconstructed 
from candidate $\Dzero$ or $\Dstarzero$ mesons that are combined with  
light neutral meson candidates $\hzero$. The $\Dzero$ mesons are reconstructed
in the final states $\Kpi$, $\Ktwopi$, and $\Kthreepi$, while the light neutral mesons  $\hzero$ are
reconstructed in the decay modes $\pizero \to \gamgam$, $\eta \to \gamgam$, $\eta \to \threepi$ 
and $\omega \to \threepi$.  
The $\Dstarzero$ mesons are reconstructed in the $\Dstarzero \to \Dzero \pizero$ decay mode.
\refmod{ The decay mode $\Bzerobar \to \Dstarzero \eta$ where $\eta \to \threepi$ is not reconstructed.}

Vertex and mass constrained fits are performed for decays with charged products such as
the  three $\Dzero$ decay modes  and $\eta \to \threepi$; mass constrained fits
are performed for 
the $\pizero \to \gamgam$ and $\eta \to \gamgam$ candidates; and vertex
constrained fits are performed for $\omega \to \threepi$ candidates. 
\mod{
These 
kinematic fits result in reduced uncertainties on the energy and momenta of the candidate mesons. 
 }
 
Charged tracks are required to have impact parameters within $\pm 5 \cm$ of the interaction
point along the 
beam axis and within $1 \cm$ in the transverse plane. Each track is identified
as a kaon or pion according to a likelihood ratio derived from the responses of the TOF and ACC
systems and energy loss measurements from the CDC. The likelihood ratio is required to exceed
0.6 for  kaon candidates; within the momentum range of interest, this requirement is $88\%$ efficient for kaons 
and has a misidentification rate for pions of $8.5\%$. 

\mod{
The photon pairs that constitute $\pizero$ candidates are required to have energies greater than
50 MeV and an invariant mass within mass windows around the nominal $\pizero$ mass
ranging between $\pm2\sigma$ to $\pm3\sigma$ depending on the $\pizero$ momentum.  
The $\pizero$ mass resolution ($\sigma$) is momentum dependent, with values in the
range $\sigma = 5.4 - 9.0 \MeVcsq$ over the momentum range of $\pizero$
produced in the $\Dstze\pizero$ decays considered.
}

Candidate $\eta$ mesons that decay to $\gamgam$ are required to have photon energies $E_{\gamma}$  greater than 
$100 \MeV$. In addition, the energy asymmetry  
$\frac{|E_{\gamma_1}-E_{\gamma_2}|}{E_{\gamma_1}+E_{\gamma_2}}$,
is required to be less than 0.9. The $\eta$ candidates are required to have invariant masses 
within a $2.5\sigma$ mass window of the nominal mass, where $\sigma = 10.6\MeVcsq$ for 
the $\eta \to \gamgam$ mode and $3.4 \MeVcsq$ for the $\eta \to \threepi$ mode.   
If either of the photons that constitute the $\eta \to \gamgam$ candidate are found to contribute to
any $\pizero \to \gamgam$ candidate, the $\eta$ decay is excluded. 
The $\pizero$ decay products 
of the $\eta \to \threepi$ and $\omega \to \threepi$ candidates 
are required to have center of mass frame (CM)  momentum greater than $200$ and $500 \MeVc$, respectively.
The $\omega$ candidates are required to have invariant masses within $\pm3 \Gamma$ of the nominal mass value, 
where $\Gamma$ is 
the natural width of \mod{ the $\omega$ meson ($\Gamma = 8.49\pm0.08 \MeVcsq$~\cite{pdg})}.

Invariant masses of the $\Dzero$ candidates are required to be within $\pm 2\sigma$ of the nominal
mass, where $\sigma$ is $8, 12$ and $5 \MeVcsq$ for the $\Kpi$, $\Ktwopi$, and $\Kthreepi$ modes, respectively.
The CM momentum of the $\pizero$ in the $\Ktwopi$ mode is required to be
greater than $400 \MeVc$.

The $\Dstarzero$ meson candidates are obtained by combining candidate $\Dzero$ and low momentum $\pizero$ mesons, 
where the soft $\pizero$ momentum \mod{in the laboratory frame} is required to be less than $0.6 \GeVc$ and the invariant mass difference 
$| M(\Dstarzero) - M(\Dzero) |$ is required to be within $2\MeVcsq$ of the nominal value.

\section{ \B reconstruction }

The $\Bzerobar$ candidates are reconstructed from combinations of  $\Dzero$ or $\Dstarzero$ and $\hzero$ using the
improved energy and momenta resulting from the vertex and mass constrained fits.  

Two kinematic variables are used to distinguish signal candidates from backgrounds: the
 beam-energy constrained mass 
 $\Mb = \sqrt{  (\Eb)^2 -  |\sum \vec{p}_{i}^{*}|^2 )}$ and the energy difference 
 $\De = \sum E_{i}^{*}  - \Eb $, 
where $\Eb$ is the CM energy, and 
$E_{i}^{*} $ , $\vec{p}_{i}^{*} $ are the CM energy and momenta, respectively, which are summed over 
the $\Dzero$ or $\Dstarzero$ and $\hzero$ meson decay candidates.

The resolution of $\Mb$ is approximately $3 \MeVcsq$ for all modes,  dominated by the  
beam energy spread, whereas the $\De$ resolution varies substantially among modes,
depending particularly on the number of $\pizero$'s in the final state.    
Candidates within the region $|\De| < 0.25 \GeV $ and $  5.2 \GeVcsq < \Mb < 5.3 \GeVcsq$
are selected for further consideration. 
Where more than one candidate or reconstruction hypothesis occurs in a single
event, a consistency measure is used to pick the best candidate and
reconstruction hypothesis. The consistency measure is
constructed from the sum of the $\chi^2$ per degree of freedom for the
relevant kinematic fits (to $\Dzero$ or $\Dstarzero$ and $\hzero$).
For hypotheses including $\Dstarzero$, an additional
term reflecting the deviation of the invariant mass difference 
$| M(\Dstarzero) - M(\Dzero) |$ from the nominal value is included in the
consistency measure. 
\refmod{
The fraction of events with multiple candidates with the same reconstruction 
hypothesis in the signal regions defined below is
estimated from signal Monte Carlo samples to be less than three percent for
all reconstructed modes.}

\mod{
Signal region definitions in $\Mb$ and $\De$ are chosen based on their resolutions.    
A common $\Mb$ signal region of  $  5.27 \GeVcsq < \Mb < 5.29 \GeVcsq$ is used for all final states. 
The signal regions in $\De$ are mode dependent, with 
$|\De| < 0.08 \GeV $ for $\Dzero\pizero$, $\Dzero\eta(\gamma\gamma)$, $\Dstarzero\pizero$ and $\Dstarzero\eta$ modes ; 
$ -0.08 \GeV < \De < 0.05 \GeV $ for $\Dzero\eta(\pizero\pi\pi)$ ;  
$ -0.12 \GeV < \De < 0.08 \GeV $ for $\Dstarzero\omega$ ; 
and 
$|\De| < 0.05 \GeV $ for $\Dzero\omega$. 
}

The event yields and efficiencies presented in the following sections correspond to these signal regions\mod{.} 
Figs.~\ref{fig:dstze_pi0},~\ref{fig:dstze_eta} and~\ref{fig:dstze_omega} show the 
$\Mb$  and $\De$ distributions after application of all selection requirements and with 
the $\De$ signal requirement applied for the $\Mb$ distributions (a,c) 
and
the signal requirement  $ 5.27 \GeVcsq  < \Mb < 5.29 \GeVcsq$ applied for the $\De$ distributions (b,d).
\mod{This selection includes continuum suppression requirements as described in the next section.}
The signal regions are indicated by vertical dashed lines on the figures.

\providecommand{\yielddecaption}{ Measured signal region yields and MC estimates of signal region contributions for $\Bzerobar \to \Dstze \hzero$ 
  for the combined $\Dzero$ subdecay modes.    
  The number of signal events ($N_{sig}$) obtained from the $\De$ fit are listed together with their statistical uncertainties. 
  For the modes $\Dstze\pizero$ estimates of the background contributions from
  $\Dstze \rho$ ($N_{D\rho}$) other $B$ backgrounds ($N_{bbk}$), continuum ($N_{q\bar{q}}$) and cross-feeds ($N_{xrs}$) are listed.   
  Combined background estimates ($N_{bkg}$) together with cross-feed contributions ($N_{xrs}$) are provided for the other $\Dzero\hzero$ modes. 
  For the other $\Dstarzero\hzero$ modes, 
  background contributions are shown for $B$ backgrounds ($N_{bbk}$), continuum ($N_{q\bar{q}}$) 
  and cross-feeds ($N_{xrs}$).  \newline\label{yield-de-dz-de-smry}
   }

\begin{table*}[htbp]
\begin{center}
\begin{center}
 \caption{\yielddecaption}
\renewcommand{\baselinestretch}{\tabbls}\small\normalsize
\begin{tabular}{@{\hspace{0.05cm}}l@{\hspace{0.05cm}}c@{\hspace{0.05cm}}c@{\hspace{0.05cm}}c@{\hspace{0.05cm}}c@{\hspace{0.05cm}}c@{\hspace{0.05cm}}c@{\hspace{0.05cm}}}\hline\hline 
{\tabelemsize{ Mode}} & {\tabelemsize{ $N_{sig}$}} & {\tabelemsize{ $N_{bbk}$}} & {\tabelemsize{ $N_{bkg}$}} & {\tabelemsize{ $N_{xrs}$}} & {\tabelemsize{ $N_{q\bar{q}}$}} & {\tabelemsize{ $N_{D\rho}$}} \\ 
\hline
{\tabelemsize{ $                          D^{0} \pi^{0} $}} & {\tabelemsize{ $  620.5 \pm    39.1$}} & {\tabelemsize{ $   26.8 \pm     2.1$}} & {\tabelemsize{ -}} & {\tabelemsize{ $   25.9 \pm     6.5$}} & {\tabelemsize{ $  448.7 \pm    28.0$}} & {\tabelemsize{ $   93.5 \pm     8.0$}} \\ 
{\tabelemsize{ $              D^{0} \eta_{\gamma\gamma} $}} & {\tabelemsize{ $  160.7 \pm    18.3$}} & {\tabelemsize{ -}} & {\tabelemsize{ $  131.8 \pm    14.6$}} & {\tabelemsize{ $    6.9 \pm     1.7$}} & {\tabelemsize{ -}} & {\tabelemsize{ -}} \\ 
{\tabelemsize{ $             D^{0} \eta_{\pi^{0}\pi\pi} $}} & {\tabelemsize{ $   64.7 \pm    11.2$}} & {\tabelemsize{ -}} & {\tabelemsize{ $   51.2 \pm     8.2$}} & {\tabelemsize{ $    2.4 \pm     0.6$}} & {\tabelemsize{ -}} & {\tabelemsize{ -}} \\ 
{\tabelemsize{ $D^{0} \eta_{\gamma\gamma + \pi^{0}\pi\pi}^{\dagger}$}}
                                                            & {\tabelemsize{ $  225.6 \pm    21.5$}} & {\tabelemsize{ -}} & {\tabelemsize{ $  174.6 \pm    16.0$}} & {\tabelemsize{ $   16.8 \pm     4.2$}} & {\tabelemsize{ -}} & {\tabelemsize{ -}} \\ 
{\tabelemsize{ $                           D^{0} \omega $}} & {\tabelemsize{ $  201.5 \pm    20.1$}} & {\tabelemsize{ -}} & {\tabelemsize{ $  135.8 \pm    12.9$}} & {\tabelemsize{ $   12.8 \pm     3.2$}} & {\tabelemsize{ -}} & {\tabelemsize{ -}} \\ 
{\tabelemsize{ $                         D^{*0} \pi^{0} $}} & {\tabelemsize{ $  115.2 \pm    14.9$}} & {\tabelemsize{ $   22.1 \pm     2.9$}} & {\tabelemsize{ -}} & {\tabelemsize{ $    2.0 \pm     0.3$}} & {\tabelemsize{ $   29.9 \pm     3.9$}} & {\tabelemsize{ $   40.5 \pm     5.3$}} \\ 
{\tabelemsize{ $             D^{*0} \eta_{\gamma\gamma} $}} & {\tabelemsize{ $   49.8 \pm    10.0$}} & {\tabelemsize{ $   19.7 \pm     3.9$}} & {\tabelemsize{ -}} & {\tabelemsize{ $    2.2 \pm     0.4$}} & {\tabelemsize{ $    8.5 \pm     1.7$}} & {\tabelemsize{ -}} \\ 
{\tabelemsize{ $                          D^{*0} \omega $}} & {\tabelemsize{ $   53.3 \pm     9.2$}} & {\tabelemsize{ $   26.4 \pm     4.5$}} & {\tabelemsize{ -}} & {\tabelemsize{ $    1.8 \pm     0.3$}} & {\tabelemsize{ $   19.5 \pm     3.4$}} & {\tabelemsize{ -}} \\ 
\hline\hline
\end{tabular}
\end{center}
\vspace*{1mm}
{\footnotesize $^\dagger$ The combined $\eta$ results are from a simultaneous fit to the individual $\eta$ samples, rather than summation of the individual yields.} 
\end{center}
\end{table*}

\section{Continuum Suppression}
At energies close to the $\Upsilon(4S)$ resonance the
production cross section of $\epem \to \qqbar$ $( q = u,d,s,c )$ is approximately three times that of $\BB$ production, 
making continuum background suppression essential in all modes. 
The jet-like nature of the continuum events allows 
event shape variables to discriminate between them and the more-spherical $\BB$ events.

Seven event-shape variables are combined into a
single Fisher discriminant~\cite{fw}. These variables 
include the angle between the thrust axis of the \B candidate and the thrust axis of the rest of the event 
($\cos{\theta_{T}}$), the sphericity variable, and five modified Fox-Wolfram
moments. \mod{The technique and details of the variables used are provided in~\cite{fw}.} 

Monte Carlo event samples of continuum $\qqbar$ events and signal events for each of the 
final states considered are used to construct probability density functions (PDFs) for the 
Fisher discriminant~\cite{fw} and $\cos{\theta_{B}}$, where $\theta_{B}$ is the angle between
the \B  flight direction and the beam direction in the 
CM
frame and additional angular variables
in some modes, as indicated below.
The products of the PDFs for these variables give signal and 
continuum likelihoods ${\cal L}_{s}$ and ${\cal L}_{\qqbar}$ for each 
candidate, allowing a selection to be applied to the likelihood ratio 
${\cal L} = {\cal L}_{s} / ({\cal L}_{s} + {\cal L}_{\qqbar} )$.    

For the decays $\Bzerobar \to \Dstarzero\pizero$ and $\Dstarzero\eta$, the vector-pseudoscalar nature of the decay products 
results in the longitudinal polarization of the $\Dstarzero$. 
The discrimination benefits from this polarization by incorporating PDFs for the $\Dstarzero$ helicity angle in the likelihood ratio. 
The $\Dstarzero$ helicity angle is defined as the angle 
between the direction of the $\Dzero$ and the opposite of the $\Bzero$
direction in the $\Dstarzero$ rest frame.
The vector-vector nature of the decay products in the 
decay  $\Bzerobar \to \Dstarzero\omega$ prevents the $\Dstarzero$ helicity
angle from being a useful discriminant in this mode,   
as the polarization of the decay products is not known. 
However, the $\omega \to \pizero \pi^{+}\pi^{-}$ ``splay'' angle, defined 
as the angle between the directions of the $\pizero$ and either the $\pi^{+}$ or $\pi^{-}$ in 
the $\pi^{+}\pi^{-}$ rest frame, is found to provide useful
discrimination and is incorporated into the likelihood ratio.  
The method used to account for uncertainties arising from the unknown
polarization is described in Section~\ref{sec:systematics}.

Monte Carlo studies of the signal significance $N_{s}/\sqrt{N_{s}+N_{b}}$,
where $N_s$ and $N_b$ are Monte Carlo signal and
background yields (using signal branching fractions from previous measurements), as a function of a cut on the
likelihood ratio ${\cal L}$ indicate a smooth behavior. Although the optimum significance is generally in the
range 0.6-0.7, a looser cut of ${\cal L} > 0.5 $  is applied for all modes in order to reduce systematic uncertainties.
This requirement removes (66--79)\% of the continuum background samples while
retaining (74--83)\% of the signal samples.

For the  $\Bzerobar \to \Dzero\omega$ mode the polarized nature of the $\omega$ allows additional discrimination against
backgrounds to be achieved with an additional requirement of $|\cos{\theta_{hel}}|>0.3$, where the
helicity angle $\theta_{hel}$  is defined as the 
angle between the \B  flight direction in the $\omega$ rest frame and the vector perpendicular to the $\omega$
decay plane in the $\omega$ rest frame.

\providecommand{\effccaption}{ 
 Efficiency correction factors for the modes $\Bzerobar \to \Dstze \hzero$; the correction factors for combined modes are averaged 
 using PDG subdecay fractions.\newline\label{effc-dz-mb-sub}
}
\begin{table*}[htbp]
\begin{center}
\begin{center}
 \caption{\effccaption}
\renewcommand{\baselinestretch}{\tabbls}\small\normalsize
\begin{tabular}{{@{\hspace{0.5cm}}l@{\hspace{0.5cm}}c@{\hspace{0.5cm}}c@{\hspace{0.5cm}}c@{\hspace{0.5cm}}c@{\hspace{0.5cm}}}}\hline\hline 
{\tabelemsize{ Mode }} & {\tabelemsize{ $                                  D^{0} $}} & {\tabelemsize{ $                            D^{0}(K\pi) $}} & {\tabelemsize{ $                     D^{0}(K\pi\pi^{0}) $}} & {\tabelemsize{ $                      D^{0}(K\pi\pi\pi) $}} \\ \hline
{\tabelemsize{ $                          D^{0} \pi^{0} $}} & {\tabelemsize{ $   0.94 \pm    0.05$}} & {\tabelemsize{ $   0.98 \pm    0.05$}} & {\tabelemsize{ $   0.90 \pm    0.04$}} & {\tabelemsize{ $   0.98 \pm    0.06$}} \\ 
{\tabelemsize{ $              D^{0} \eta_{\gamma\gamma} $}} & {\tabelemsize{ $   0.99 \pm    0.06$}} & {\tabelemsize{ $   1.03 \pm    0.06$}} & {\tabelemsize{ $   0.95 \pm    0.05$}} & {\tabelemsize{ $   1.03 \pm    0.07$}} \\ 
{\tabelemsize{ $             D^{0} \eta_{\pi^{0}\pi\pi} $}} & {\tabelemsize{ $   0.94 \pm    0.06$}} & {\tabelemsize{ $   0.98 \pm    0.05$}} & {\tabelemsize{ $   0.91 \pm    0.05$}} & {\tabelemsize{ $   0.98 \pm    0.06$}} \\ 
{\tabelemsize{ $D^{0} \eta_{\gamma\gamma + \pi^{0}\pi\pi}  $}} & {\tabelemsize{ $   0.97 \pm    0.06$}} &  - & - & - \\ 
{\tabelemsize{ $                           D^{0} \omega $}} & {\tabelemsize{ $   0.89 \pm    0.06$}} & {\tabelemsize{ $   0.92 \pm    0.06$}} & {\tabelemsize{ $   0.85 \pm    0.06$}} & {\tabelemsize{ $   0.92 \pm    0.07$}} \\ 
{\tabelemsize{ $                         D^{*0} \pi^{0} $}} & {\tabelemsize{ $   0.86 \pm    0.09$}} & {\tabelemsize{ $   0.90 \pm    0.10$}} & {\tabelemsize{ $   0.83 \pm    0.09$}} & {\tabelemsize{ $   0.90 \pm    0.10$}} \\ 
{\tabelemsize{ $             D^{*0} \eta_{\gamma\gamma} $}} & {\tabelemsize{ $   0.91 \pm    0.10$}} & {\tabelemsize{ $   0.95 \pm    0.10$}} & {\tabelemsize{ $   0.88 \pm    0.10$}} & {\tabelemsize{ $   0.95 \pm    0.11$}} \\ 
{\tabelemsize{ $                          D^{*0} \omega $}} & {\tabelemsize{ $   0.83 \pm    0.10$}} & {\tabelemsize{ $   0.87 \pm    0.10$}} & {\tabelemsize{ $   0.80 \pm    0.09$}} & {\tabelemsize{ $   0.88 \pm    0.11$}} \\ 
\hline\hline
\end{tabular}
\end{center}
\end{center}
\end{table*}

\providecommand{\coreffcaption}{ 
Corrected efficiencies for the modes $\Bzerobar \to \Dstze \hzero$ excluding the subdecay branching fractions,
for all $\Dzero$ modes combined and for the individual $\Dzero$ subdecay modes,  
as estimated for the $\De$  fit samples.   \newline\label{efxvnim-dz-de-sub}}
\begin{table*}[htbp]
\begin{center}
\begin{center}
 \caption{\coreffcaption}
\renewcommand{\baselinestretch}{\tabbls}\small\normalsize
\begin{tabular}{{@{\hspace{0.5cm}}l@{\hspace{0.5cm}}c@{\hspace{0.5cm}}c@{\hspace{0.5cm}}c@{\hspace{0.5cm}}c@{\hspace{0.5cm}}}}\hline\hline 
{\tabelemsize{ Mode}} & {\tabelemsize{ $                                  D^{0} $}} & {\tabelemsize{ $                            D^{0}(K\pi) $}} & {\tabelemsize{ $                     D^{0}(K\pi\pi^{0}) $}} & {\tabelemsize{ $                      D^{0}(K\pi\pi\pi) $}} \\ \hline
{\tabelemsize{ $                          D^{0} \pi^{0} $}} & {\tabelemsize{ $  0.075 \pm   0.004$}} & {\tabelemsize{ $  0.173 \pm   0.008$}} & {\tabelemsize{ $  0.042 \pm   0.002$}} & {\tabelemsize{ $  0.085 \pm   0.005$}} \\ 
{\tabelemsize{ $              D^{0} \eta_{\gamma\gamma} $}} & {\tabelemsize{ $  0.061 \pm   0.004$}} & {\tabelemsize{ $  0.140 \pm   0.008$}} & {\tabelemsize{ $  0.036 \pm   0.002$}} & {\tabelemsize{ $  0.066 \pm   0.004$}} \\ 
{\tabelemsize{ $             D^{0} \eta_{\pi^{0}\pi\pi} $}} & {\tabelemsize{ $  0.042 \pm   0.003$}} & {\tabelemsize{ $  0.092 \pm   0.005$}} & {\tabelemsize{ $  0.026 \pm   0.002$}} & {\tabelemsize{ $  0.046 \pm   0.003$}} \\ 
{\tabelemsize{ $D^{0} \eta_{\gamma\gamma + \pi^{0}\pi\pi}  $}} & {\tabelemsize{ $  0.054 \pm   0.003$}} & {\tabelemsize{ $  0.122 \pm   0.010$}} & {\tabelemsize{ $  0.032 \pm   0.003$}} & {\tabelemsize{ $  0.058 \pm   0.006$}} \\ 
{\tabelemsize{ $                           D^{0} \omega $}} & {\tabelemsize{ $  0.025 \pm   0.002$}} & {\tabelemsize{ $  0.056 \pm   0.004$}} & {\tabelemsize{ $  0.015 \pm   0.001$}} & {\tabelemsize{ $  0.030 \pm   0.002$}} \\ 
{\tabelemsize{ $                         D^{*0} \pi^{0} $}} & {\tabelemsize{ $  0.036 \pm   0.004$}} & {\tabelemsize{ $  0.088 \pm   0.009$}} & {\tabelemsize{ $  0.019 \pm   0.002$}} & {\tabelemsize{ $  0.042 \pm   0.005$}} \\ 
{\tabelemsize{ $             D^{*0} \eta_{\gamma\gamma} $}} & {\tabelemsize{ $  0.039 \pm   0.004$}} & {\tabelemsize{ $  0.093 \pm   0.010$}} & {\tabelemsize{ $  0.023 \pm   0.003$}} & {\tabelemsize{ $  0.042 \pm   0.005$}} \\ 
{\tabelemsize{ $                          D^{*0} \omega $}} & {\tabelemsize{ $  0.011 \pm   0.001$}} & {\tabelemsize{ $  0.026 \pm   0.003$}} & {\tabelemsize{ $  0.007 \pm   0.001$}} & {\tabelemsize{ $  0.011 \pm   0.001$}} \\ 
\hline\hline
\end{tabular}
\end{center}
\end{center}
\end{table*}

\section{Backgrounds from other \B decays}
Significant background contributions arise both from
color-favored decays 
and from other color-suppressed decays (cross-feed) 
$\Bzerobar \to \Dstarzero \hzero $.  
Some backgrounds have the same final state
as the signal while others mimic signal due to missing or extra particles.

Generic Monte Carlo~\cite{ref:bellemc} samples of $\BB$ and continuum $\qqbar$
are used to study the background contributions in the $\Mb$ and $\De$ distributions.
\refmod{The sample sizes correspond to approximately three times the
expectations from the data sample analysed.}
The $\BB$ event sample excludes the color-suppressed modes under investigation. 
\mod{Signal mode samples for each of the decay chains considered are
generated and reconstructed separately.}
They are used to estimate the cross-feed contributions between modes using the branching 
fractions measured here \mod{with an iterative procedure}.

The dominant cross-feed contributions to the $\Dzero \hzero$ decays are found to arise from the corresponding
$\Dstarzero \hzero$ decays. These contributions peak at the same $\Mb$ as the signal but are shifted to the lower side
in $\De$. 
\mod{
As can be seen from Figs.~\ref{fig:dstze_eta}b and ~\ref{fig:dstze_omega}b,  
the cross-feed contribution is substantial in the region $ -0.25 \GeV < \De < -0.10 \GeV$
but quite small in the signal region.    
}
Cross-feed contributions to the $\Dstarzero \hzero$ decays are found to be small.
In all cases, the fraction of cross-feed within the                                                                          
signal region is less than 10\% of the observed yield.

For the $\Bzerobar \to \Dstze\pizero$ modes, the dominant background is from 
the color-allowed $\Bmi \to \Dstze \rhomi  $ decay.
In the $\Bzerobar \to \Dzero\pizero$ mode  non-reconstructed soft $\pizero$ from $\Dstarzero \to \Dzero \pizero $, 
photons from $\Dstarzero \to \Dzero \gamma $ and $\pimi$ 
from  $ \rhomi \to \pimi \pizero $  produce the same final state as the signal. 
However, the missing particles cause a shift in $\De$ with a broad peak centered at approximately $\De = -0.2 \GeV$. 
In order to reduce contributions from this background, events that contain \B candidates
reconstructed as   $\Bmi \to \Dstze \rhomi $  within the signal region $ 5.27 \GeV < \Mb < 5.29 \GeV$ and $|\De| < 0.1 \GeV $ are rejected.
This requirement reduces the color-allowed contribution in the region $-0.25 \GeV < \De < -0.10 \GeV$  by about $60 \%$; 
it does little to reduce contributions in the signal region, but remains useful to facilitate background modelling. 
The $\Mb$ distribution of these backgrounds is found to contribute at and slightly below the $\Mb$ signal region; these are
referred to as the ``peaking background''.

For the $\Bzerobar$ decays to $\Dstze\eta$ and $\Dstze\omega$ modes there are
potential backgrounds arising from non-resonant $\Bzerobar \to \Dstze \pi^{+}\pi^{-}\pi^{0}$ decays. 
Invariant mass, $M(\pi^{+}\pi^{-}\pi^{0})$, distributions within the $\Mb$ and $\De$
signal regions indicate no significant contributions from these non-resonant decays.
\refmod{Ratios of data to Monte Carlo expectations in the invariant mass sidebands 
give values consistent with 1.0 with relative uncertainties in the range of $3$--$8$ percent.
As there are no indications of the presence of this background no systematic uncertainties 
from this source are assigned.}

\begin{table*}[htbp]
\begin{center}
\begin{center}
\caption{  Measured branching fractions $(\times 10^{-4})$ for the modes $\Bzerobar \to \Dstze \hzero$,  using separate $\Dzero$ subdecay mode samples,   as obtained from the $\De$ fit.   \newline\label{xbrrnim-dz-de-sub}}\renewcommand{\baselinestretch}{\tabbls}\small\normalsize
\begin{tabular}{{@{\hspace{0.5cm}}l@{\hspace{0.5cm}}c@{\hspace{0.5cm}}c@{\hspace{0.5cm}}c@{\hspace{0.5cm}}}}\hline\hline 
{\tabelemsize{ Mode }} & {\tabelemsize{ $                            D^{0}(K\pi) $}} & {\tabelemsize{ $                     D^{0}(K\pi\pi^{0}) $}} & {\tabelemsize{ $                      D^{0}(K\pi\pi\pi) $}} \\ \hline
{\tabelemsize{ $                          D^{0} \pi^{0} $}} & {\tabelemsize{ $   2.13 \pm    0.19 \pm    0.31 $}} & {\tabelemsize{ $   2.02 \pm    0.24 \pm    0.33 $}} & {\tabelemsize{ $   2.43 \pm    0.27 \pm    0.38 $}} \\ 
{\tabelemsize{ $              D^{0} \eta_{\gamma\gamma} $}} & {\tabelemsize{ $   1.91 \pm    0.29 \pm    0.24 $}} & {\tabelemsize{ $   1.64 \pm    0.32 \pm    0.24 $}} & {\tabelemsize{ $   1.66 \pm    0.40 \pm    0.23 $}} \\ 
{\tabelemsize{ $             D^{0} \eta_{\pi^{0}\pi\pi} $}} & {\tabelemsize{ $   1.52 \pm    0.44 \pm    0.26 $}} & {\tabelemsize{ $   1.51 \pm    0.46 \pm    0.28 $}} & {\tabelemsize{ $   2.18 \pm    0.60 \pm    0.39 $}} \\ 
{\tabelemsize{ $                           D^{0} \omega $}} & {\tabelemsize{ $   2.57 \pm    0.37 \pm    0.27 $}} & {\tabelemsize{ $   1.77 \pm    0.36 \pm    0.22 $}} & {\tabelemsize{ $   2.60 \pm    0.42 \pm    0.31 $}} \\ 
{\tabelemsize{ $                         D^{*0} \pi^{0} $}} & {\tabelemsize{ $   1.21 \pm    0.23 \pm    0.17 $}} & {\tabelemsize{ $   1.45 \pm    0.36 \pm    0.23 $}} & {\tabelemsize{ $   1.54 \pm    0.38 \pm    0.23 $}} \\ 
{\tabelemsize{ $             D^{*0} \eta_{\gamma\gamma} $}} & {\tabelemsize{ $   1.43 \pm    0.42 \pm    0.25 $}} & {\tabelemsize{ $   0.79 \pm    0.44 \pm    0.15 $}} & {\tabelemsize{ $   2.09 \pm    0.57 \pm    0.38 $}} \\ 
{\tabelemsize{ $                          D^{*0} \omega $}} & {\tabelemsize{ $   2.66 \pm    0.64 \pm    0.38 $}} & {\tabelemsize{ $   1.90 \pm    0.62 \pm    0.31 $}} & {\tabelemsize{ $   2.12 \pm    0.76 \pm    0.38 $}} \\ 
\hline\hline
\end{tabular}
\end{center}
\end{center}
\end{table*}
\section{Data modelling and Signal Extraction}
Independent unbinned extended maximum likelihood fits to the $\De$ and $\Mb$ distributions are performed to obtain the signal yields. 
The yields from the $\De$ fits are used to extract the branching fractions, while the yields from the $\Mb$ 
fits are used to cross-check the results. The principal parameters of the fits are the normalization factors of the components used
to model the observed distributions. 
\mod{
The $\De$ fit is performed in the range $ -0.25 \GeV < \De < 0.25 \GeV$ 
using the $\Mb$ signal sample while the 
$\Mb$ fit 
is performed in the range $\Mb > 5.2 \GeVcsq$ 
using the mode-dependent $\De$ signal samples. 
}
In most cases the shapes of the signal and background component distributions in $\Mb$ and $\De$ are obtained 
from fits to MC samples.

The signal models used are the same for all modes,  with the $\Mb$ signals modeled with a Gaussian function
 and the $\De$ signals modeled with an empirical formula known as the Crystal
 Ball (CB) line shape~\cite{cbline}, that accounts for the asymmetric
 calorimeter energy response. The CB function is added to a Gaussian function \refmod{ in $\De$ } with the same mean. 
Fits of the $\De$ distributions to signal Monte Carlo for each final state are used to obtain the 
signal shape parameters.

 The cross-feed contributions in $\Mb$ and $\De$ are studied using a combination of signal Monte Carlo samples from
 all other color suppressed modes, \refmod{weighted according to the branching fractions obtained here.}
 Smoothed histograms obtained from this combined sample are used as estimates of the  cross-feed 
 contributions.  In the $\Mb$ case the $\De$ signal region requirement results in very small cross-feed contributions, which
 are fixed at the Monte Carlo expectation. For $\De$ there are considerable contributions 
in the region  $ -0.25 \GeV < \De < -0.10 \GeV$. 
The normalization of this component is allowed to float in the fit for the
$\Dzero \hzero$ modes; for the $\Dstarzero \hzero$ modes the cross-feed
expectations are small and are fixed at the Monte Carlo expectation.

 Continuum-like backgrounds in the $\Mb$ fits are modeled by an empirical threshold function known as 
 the ARGUS function~\cite{argus}.  The small peaking background contributions 
 are modeled by a Gaussian of mean and width and normalization obtained by a fit to the $B\bar{B}$ background Monte Carlo $\Mb$      distribution,  using an ARGUS function plus a Gaussian. A systematic uncertainty of 50\% is assigned to the determination
 of this small background distribution.   
 This treatment allows the vast majority of the background to be simply
 modeled with the ARGUS shape, leaving a 
 small but less well-known peaking background component  that represents the deviation from the ARGUS shape. 
 
 In fits to data $\Mb$ distributions, the ARGUS background function parameters 
 are fixed to the values obtained from fits to combined Monte Carlo
 $B\bar{B}$ and continuum background samples.  The signal parameters are free, as are the 
 normalizations of signal and background. 
 The small peaking background and cross-feed contributions are fixed at their expected values.
  
  
The $\De$ background distributions in the $\Bzerobar \to \Dzero\eta$ 
and  $\Bzerobar \to \Dzero\omega$ are modeled  using smoothed histograms obtained from a combined continuum and
generic $\BB$ Monte Carlo sample.   
For the $\Bzerobar \to \Dstze\pizero$ modes, the shapes of the $\De$ distribution arising 
from $\BB$ and continuum background are very different (see Fig.~\ref{fig:dstze_pi0}) necessitating separate modelling. 
The continuum shape is modeled with a first-order polynomial with slope obtained from fits to the continuum
Monte Carlo sample. The shape of the $\BB$ background is modeled with   
a Gaussian function plus a second-order polynomial, with parameters determined from a fit to  
the generic $\BB$ Monte Carlo sample. In fits to data, the large peak in the region $ -0.25 \GeV < \De < -0.10 \GeV $ that
arises principally from the color-allowed $\Bmi \to \Dstze \rhomi $ decays is found to be broader than the Monte Carlo
expectation; thus all parameters of this color-allowed Gaussian are allowed to float in the fit. The normalizations of the contributions 
from the remainder of the $\BB$ background, the continuum and the signal are also floated in the fit, with the small cross-feed contribution
fixed as discussed above. 

The results of the $\Mb$ and $\De$ fits for each of the $\Dstze\hzero$ modes with the three $\Dzero$ subdecay modes combined
are presented in Figs.~\ref{fig:dstze_pi0}, ~\ref{fig:dstze_eta} and ~\ref{fig:dstze_omega}.
The $\Dstze\hzero$ mode results are obtained by a simultaneous fit to the
three submodes, where the signal yields in each submode are constrained by the
ratios of the products of efficiency and secondary branching fraction.

\providecommand{\syscaption}{ 
Systematic uncertainties of the measured branching fractions for $\Bzerobar
\to \Dstze \hzero $, for the combined $\Dzero$ submode samples,   as estimated
for the $\De$ fit results.  \newline\label{stotxim-dz-de-smry}}
\begin{table*}[htbp]
\begin{center}
\begin{center}
 \caption{\syscaption}
\renewcommand{\baselinestretch}{\tabbls}\small\normalsize
\begin{tabular}{@{\hspace{0.05cm}}l@{\hspace{0.05cm}}c@{\hspace{0.05cm}}c@{\hspace{0.05cm}}c@{\hspace{0.05cm}}c@{\hspace{0.05cm}}c@{\hspace{0.05cm}}c@{\hspace{0.05cm}}c@{\hspace{0.05cm}}c@{\hspace{0.05cm}}}\hline\hline 
{\tabelemsize{ Category}} & {\tabelemsize{ $                          D^{0} \pi^{0} $}} & {\tabelemsize{ $              D^{0} \eta_{\gamma\gamma} $}} & {\tabelemsize{ $             D^{0} \eta_{\pi^{0}\pi\pi} $}} & {\tabelemsize{ $D^{0} \eta_{\gamma\gamma + \pi^{0}\pi\pi}  $}} & {\tabelemsize{ $                           D^{0} \omega $}} & {\tabelemsize{ $                         D^{*0} \pi^{0} $}} & {\tabelemsize{ $             D^{*0} \eta_{\gamma\gamma} $}} & {\tabelemsize{ $                          D^{*0} \omega $}} \\ 
\hline
{\tabelemsize{ Tracking efficiency}} & {\tabelemsize{ 2.6}} & {\tabelemsize{ 2.6}} & {\tabelemsize{ 2.6}} & {\tabelemsize{ 2.6}} & {\tabelemsize{ 2.6}} & {\tabelemsize{ 2.6}} & {\tabelemsize{ 2.6}} & {\tabelemsize{ 2.6}} \\ 
{\tabelemsize{ $\hzero$ efficiency}} & {\tabelemsize{ 2.7}} & {\tabelemsize{ 4.0}} & {\tabelemsize{ 4.0}} & {\tabelemsize{ 4.0}} & {\tabelemsize{ 5.4}} & {\tabelemsize{ 2.7}} & {\tabelemsize{ 4.0}} & {\tabelemsize{ 5.4}} \\ 
{\tabelemsize{ Kaon efficiency}} & {\tabelemsize{ 1.0}} & {\tabelemsize{ 1.0}} & {\tabelemsize{ 1.0}} & {\tabelemsize{ 1.0}} & {\tabelemsize{ 1.0}} & {\tabelemsize{ 1.0}} & {\tabelemsize{ 1.0}} & {\tabelemsize{ 1.0}} \\ 
{\tabelemsize{ Extra $\pizero$ efficiency }} & {\tabelemsize{ 0.8}} & {\tabelemsize{ 0.8}} & {\tabelemsize{ 0.8}} & {\tabelemsize{ 0.8}} & {\tabelemsize{ 0.8}} & {\tabelemsize{ 0.8}} & {\tabelemsize{ 0.8}} & {\tabelemsize{ 0.8}} \\ 
{\tabelemsize{ Likelihood ratio efficiency }} & {\tabelemsize{ 3.0}} & {\tabelemsize{ 3.0}} & {\tabelemsize{ 3.0}} & {\tabelemsize{ 3.0}} & {\tabelemsize{ 3.0}} & {\tabelemsize{ 3.0}} & {\tabelemsize{ 3.0}} & {\tabelemsize{ 3.0}} \\ 
{\tabelemsize{ MC statistics  }} & {\tabelemsize{ 2.1}} & {\tabelemsize{ 2.1}} & {\tabelemsize{ 2.6}} & {\tabelemsize{ 2.3}} & {\tabelemsize{ 3.0}} & {\tabelemsize{ 3.3}} & {\tabelemsize{ 2.6}} & {\tabelemsize{ 5.3}} \\ 
{\tabelemsize{ Slow $\pizero$ (from $\Dstarzero$), efficiency }} & {\tabelemsize{ 0.0}} & {\tabelemsize{ 0.0}} & {\tabelemsize{ 0.0}} & {\tabelemsize{ 0.0}} & {\tabelemsize{ 0.0}} & {\tabelemsize{ 9.5}} & {\tabelemsize{ 9.5}} & {\tabelemsize{ 9.5}} \\ 
{\tabelemsize{ Crossfeed}} & {\tabelemsize{ 1.1}} & {\tabelemsize{ 1.1}} & {\tabelemsize{ 1.1}} & {\tabelemsize{ 1.1}} & {\tabelemsize{ 1.6}} & {\tabelemsize{ 0.4}} & {\tabelemsize{ 1.1}} & {\tabelemsize{ 0.9}} \\ 
{\tabelemsize{ Modelling $\pm1\sigma$ variations}} & {\tabelemsize{ 13.6}} & {\tabelemsize{ 11.1}} & {\tabelemsize{ 16.0}} & {\tabelemsize{ 8.3}} & {\tabelemsize{ 7.4}} & {\tabelemsize{ 12.9}} & {\tabelemsize{ 12.3}} & {\tabelemsize{ 6.4}} \\ 
{\tabelemsize{ Subdecay Branching Fractions }} & {\tabelemsize{ 5.2}} & {\tabelemsize{ 5.2}} & {\tabelemsize{ 5.5}} & {\tabelemsize{ 5.3}} & {\tabelemsize{ 5.2}} & {\tabelemsize{ 7.1}} & {\tabelemsize{ 7.1}} & {\tabelemsize{ 7.1}} \\ 
{\tabelemsize{ Number of $B\bar{B}$ events}} & {\tabelemsize{ 0.7}} & {\tabelemsize{ 0.7}} & {\tabelemsize{ 0.7}} & {\tabelemsize{ 0.7}} & {\tabelemsize{ 0.7}} & {\tabelemsize{ 0.7}} & {\tabelemsize{ 0.7}} & {\tabelemsize{ 0.7}} \\ 
{\tabelemsize{ Longitudinal polarization fraction}} & {\tabelemsize{ -}} & {\tabelemsize{ -}} & {\tabelemsize{ -}} & {\tabelemsize{ -}} & {\tabelemsize{ -}} & {\tabelemsize{ -}} & {\tabelemsize{ -}} & {\tabelemsize{ 7.1}} \\ 
\hline
{\tabelemsize{ Total (\%)}} & {\tabelemsize{ 15.6}} & {\tabelemsize{ 13.8}} & {\tabelemsize{ 18.1}} & {\tabelemsize{ 11.7}} & {\tabelemsize{ 11.9}} & {\tabelemsize{ 18.5}} & {\tabelemsize{ 18.3}} & {\tabelemsize{ 17.6}} \\ 
\hline\hline
\end{tabular}
\end{center}
\end{center}
\end{table*}
\section{Branching Fraction results} 
\mod{The yields obtained} from the $\Mb$ and $\De$ fits are consistent; the
difference is typically within 50\% of the statistical uncertainty. 
The results from the $\De$ fits are found to have a slightly smaller total uncertainty in most cases and are used for the final result.
The yields for the $\Dstze\hzero$ modes (with the three $\Dzero$ subdecay samples combined) obtained from  
the one dimensional $\De$ fits are presented in Table~\ref{yield-de-dz-de-smry}. 
The statistical significances of the signals for each of the combined subdecay
samples are greater than $6\sigma$.
\mod{The combined $\eta \to \gamgam$ and  $\eta \to \threepi$ yields are
obtained from a combined fit to the individual $\eta$ samples, rather
than from a summation of the individual sample yields. The agreement between these
approaches is apparent from the Table.
}
For the $\Dzero\hzero$ modes, backgrounds arising from cross-feed 
contributions in the signal region can be seen to contribute substantially less than the 
extent of the statistical uncertainty on the signal yield.

The yields obtained are interpreted as branching fractions using the number of $\BB$ events,
the product of subdecay fractions~\cite{pdg}  corresponding to the decay of  $\Dzero \hzero$ or $\Dstarzero\hzero$ into the 
observed final states, and the efficiency. The efficiency for each mode is first obtained from signal Monte Carlo
samples and then corrected 
\mod{ to account for differences between data and MC expectations.}

The total efficiency corrections for each final state are obtained from the
product of the relevant efficiency correction factors. The values for
all final states of the $\Dstze\hzero$ modes are presented in Table~\ref{effc-dz-mb-sub} 
and the 
efficiencies are presented in Table~\ref{efxvnim-dz-de-sub}. 
 
For the $\Dstarzero\hzero$ modes, the correction to the reconstruction
efficiency of the soft pion   
produced in the process $\Dstarzero \to \Dzero \pizero $ is estimated to be $0.920 \pm 0.087$; this 
dominates the correction value and uncertainty for these modes.
This soft pion efficiency correction is estimated by comparing the yields of $\Dstarpl$ from the processes 
$\Dstarpl \to \Dzero\pipl$ and $\Dstarpl \to \Dpl\pizero$, using the subdecay modes 
$\Dzero \to \Kpi$, $\Dzero \to \Kthreepi$ and $\Dpl \to \Ktwopipl$. 
The final states differ by a single $\pizero$ or $\pipl$, allowing the ratio of yields
of $\Dstarpl \to \Dzero\pipl$ and $\Dstarpl \to \Dpl\pizero$ to be used to estimate the ratio of efficiencies 
between soft $\pipl$ and $\pizero$.  Forming a double ratio of data over Monte Carlo and using 
separate estimates for the soft $\pipl$ efficiency correction provides the soft $\pizero$ efficiency correction factor. 
The uncertainties on this factor are dominated by the uncertainties on the subdecay branching fractions.    
Reconstruction efficiencies for more energetic $\pizero$'s are obtained from comparisons 
of $\eta \to \pizero \pizero \pizero $ to $\eta \to \gamgam$ and to $\eta \to \threepi$, for data and Monte Carlo.

\begin{table}[htb]
\begin{center}
\begin{center}
\caption{  Measured branching fractions $(\times 10^{-4})$ for the modes $\Bzerobar \to \Dstze \hzero$,  obtained from simultaneous fits to the three $\Dzero$ subdecay mode samples   as obtained from the $\De$ fit.   \newline\label{xbrrsim-dz-de-smry}}\renewcommand{\baselinestretch}{\tabbls}\small\normalsize
\begin{tabular}{{@{\hspace{0.5cm}}l@{\hspace{0.5cm}}c@{\hspace{0.5cm}}}}\hline\hline 
{\tabelemsize{ Mode }} & {\tabelemsize{ Branching fraction $( \times 10^{-4} )$}} \\ \hline
{\tabelemsize{ $                          D^{0} \pi^{0} $}} & {\tabelemsize{ $   2.25 \pm    0.14 \pm    0.35 $}} \\ 
{\tabelemsize{ $              D^{0} \eta_{\gamma\gamma} $}} & {\tabelemsize{ $   1.77 \pm    0.20 \pm    0.24 $}} \\ 
{\tabelemsize{ $             D^{0} \eta_{\pi^{0}\pi\pi} $}} & {\tabelemsize{ $   1.78 \pm    0.30 \pm    0.32 $}} \\ 
{\tabelemsize{ $D^{0} \eta_{\gamma\gamma + \pi^{0}\pi\pi}  $}} & {\tabelemsize{ $   1.77 \pm    0.16 \pm    0.21 $}} \\ 
{\tabelemsize{ $                           D^{0} \omega $}} & {\tabelemsize{ $   2.37 \pm    0.23 \pm    0.28 $}} \\ 
{\tabelemsize{ $                         D^{*0} \pi^{0} $}} & {\tabelemsize{ $   1.39 \pm    0.18 \pm    0.26 $}} \\ 
{\tabelemsize{ $             D^{*0} \eta_{\gamma\gamma} $}} & {\tabelemsize{ $   1.40 \pm    0.28 \pm    0.26 $}} \\ 
{\tabelemsize{ $                          D^{*0} \omega $}} & {\tabelemsize{ $   2.29 \pm    0.39 \pm    0.40 $}} \\ 
\hline\hline
\end{tabular}
\end{center}
\end{center}
\end{table}
\begin{table}[htb]
\begin{center}
\begin{center}
\caption{Ratios of branching fractions, $\frac{ \Bzerobar \ra \Dzero \hzero }{ \Bzerobar \ra \Dstarzero \hzero }$,  with statistical and systematic uncertainties. \label{xbrrx_01}}
\renewcommand{\baselinestretch}{\tabbls}\small\normalsize
\begin{tabular}{lc}\hline\hline 
{\tabelemsize{ Modes}} & {\tabelemsize{ Ratio of branching fractions }} \\ \hline
{\tabelemsize{ ${                          D^{0} \pi^{0} } \over {                         D^{*0} \pi^{0} }$}} & {\tabelemsize{ $   1.62 \pm    0.23 \pm    0.35 $}} \\ 
{\tabelemsize{ ${              D^{0} \eta_{\gamma\gamma} } \over {             D^{*0} \eta_{\gamma\gamma} }$}} & {\tabelemsize{ $   1.27 \pm    0.29 \pm    0.25 $}} \\ 
{\tabelemsize{ ${                           D^{0} \omega } \over {                          D^{*0} \omega }$}} & {\tabelemsize{ $   1.04 \pm    0.20 \pm    0.17 $}} \\ 
\hline\hline
\end{tabular}
\end{center}
\end{center}
\end{table}

\begin{figure}
\begin{tabular}{c@{}c} 
\includegraphics[width=0.25\textwidth]{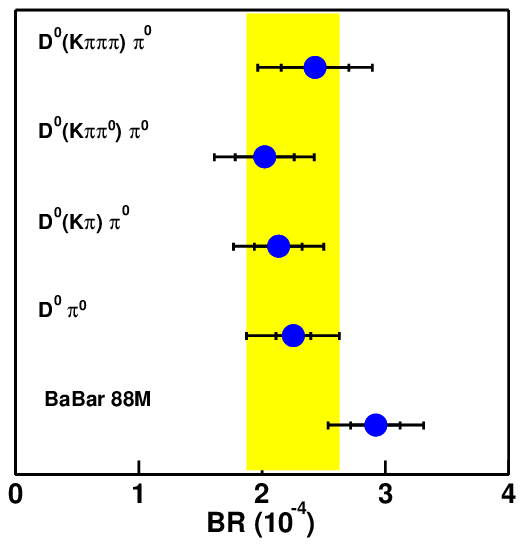}
 & 
\includegraphics[width=0.25\textwidth]{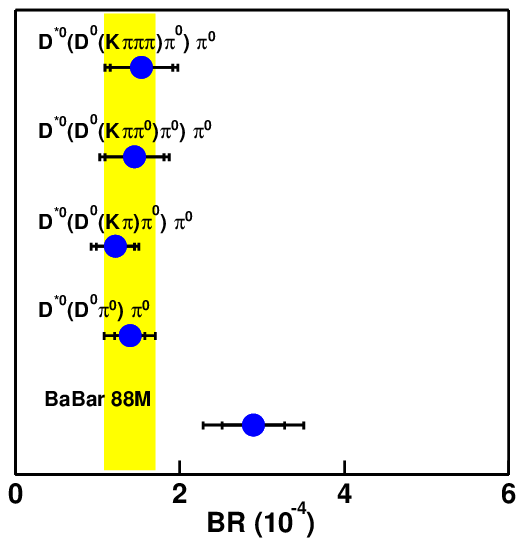}
  
\hspace{-0.25cm}
\\ 
\includegraphics[width=0.25\textwidth]{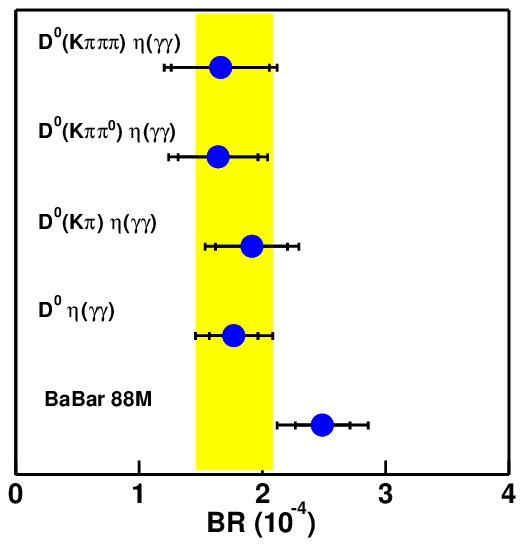}
 & 
\includegraphics[width=0.25\textwidth]{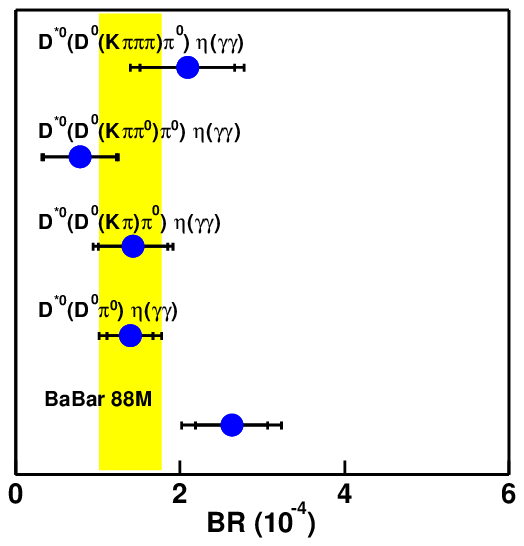}

\hspace{-0.25cm}
\\ 
\includegraphics[width=0.25\textwidth]{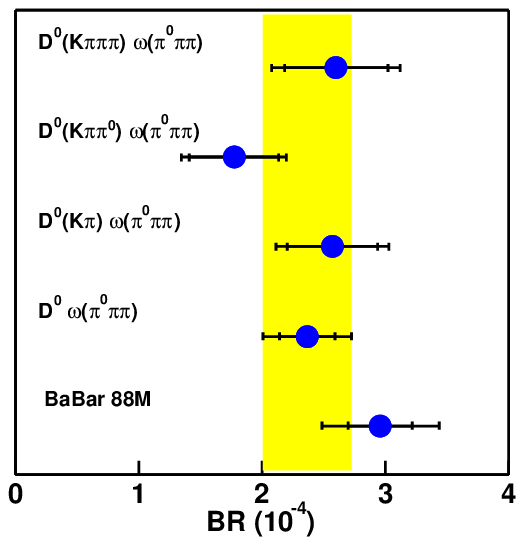}
 & 
\includegraphics[width=0.25\textwidth]{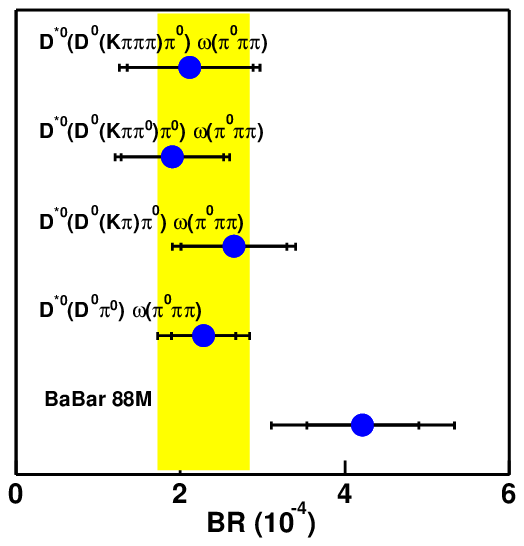}

\hspace{-0.25cm}
\end{tabular}
\begin{center}
\caption[]{
\cffigcaption
\label{fig:dstze_cf}
}
\end{center}
\end{figure}

The branching fraction results obtained from the $\De$ fit yields  
for the $\Dstze\hzero$ modes with the $\Dzero$ submodes combined are provided in 
Table~\ref{xbrrsim-dz-de-smry}. Results from the individual submodes are shown in Table~\ref{xbrrnim-dz-de-sub}. 
These results are also graphically presented in Fig.~\ref{fig:dstze_cf}, where the BaBar~\cite{ref:Babar} results
from a one dimensional fit to $\Mb$ distributions are also included for comparison. 
\mod{
Table~\ref{xbrrx_01}
presents the measured ratios of branching fractions, which benefit from a
partial cancellation of systematic uncertainties.  
}

\providecommand{\dependencycaption}{
Dependency coefficients expressing the relative changes to the 
branching fractions provided in Table~\ref{xbrrsim-dz-de-smry} 
due to relative changes to the $\Dzero$ branching fractions. 
The uncertainties presented are propagated from the uncertainties on the efficiencies only.\newline\label{ffxvnim-dz-de-sub}
}
\begin{table}[htbp]
\begin{center}
\begin{center}
 \caption{\dependencycaption}
\renewcommand{\baselinestretch}{\tabbls}\small\normalsize
\begin{tabular}{{@{\hspace{0.5cm}}l@{\hspace{0.5cm}}c@{\hspace{0.5cm}}c@{\hspace{0.5cm}}c@{\hspace{0.5cm}}}}\hline\hline 
{\tabelemsize{ Mode}} & {\tabelemsize{ $                            D^{0}(K\pi) $}} & {\tabelemsize{ $                     D^{0}(K\pi\pi^{0}) $}} & {\tabelemsize{ $                      D^{0}(K\pi\pi\pi) $}} \\ \hline
{\tabelemsize{ $                          D^{0} \pi^{0} $}} & {\tabelemsize{ $  0.352 \pm   0.034$}} & {\tabelemsize{ $  0.311 \pm   0.030$}} & {\tabelemsize{ $  0.337 \pm   0.032$}} \\ 
{\tabelemsize{ $              D^{0} \eta_{\gamma\gamma} $}} & {\tabelemsize{ $  0.350 \pm   0.038$}} & {\tabelemsize{ $  0.329 \pm   0.036$}} & {\tabelemsize{ $  0.321 \pm   0.035$}} \\ 
{\tabelemsize{ $             D^{0} \eta_{\pi^{0}\pi\pi} $}} & {\tabelemsize{ $  0.335 \pm   0.037$}} & {\tabelemsize{ $  0.339 \pm   0.038$}} & {\tabelemsize{ $  0.327 \pm   0.036$}} \\ 
{\tabelemsize{ $D^{0} \eta_{\gamma\gamma + \pi^{0}\pi\pi}  $}} & {\tabelemsize{ $  0.344 \pm   0.038$}} & {\tabelemsize{ $  0.333 \pm   0.036$}} & {\tabelemsize{ $  0.323 \pm   0.035$}} \\ 
{\tabelemsize{ $                           D^{0} \omega $}} & {\tabelemsize{ $  0.334 \pm   0.044$}} & {\tabelemsize{ $  0.318 \pm   0.042$}} & {\tabelemsize{ $  0.347 \pm   0.045$}} \\ 
{\tabelemsize{ $                         D^{*0} \pi^{0} $}} & {\tabelemsize{ $  0.368 \pm   0.072$}} & {\tabelemsize{ $  0.285 \pm   0.056$}} & {\tabelemsize{ $  0.346 \pm   0.068$}} \\ 
{\tabelemsize{ $             D^{*0} \eta_{\gamma\gamma} $}} & {\tabelemsize{ $  0.362 \pm   0.072$}} & {\tabelemsize{ $  0.319 \pm   0.064$}} & {\tabelemsize{ $  0.319 \pm   0.064$}} \\ 
{\tabelemsize{ $                          D^{*0} \omega $}} & {\tabelemsize{ $  0.356 \pm   0.079$}} & {\tabelemsize{ $  0.341 \pm   0.076$}} & {\tabelemsize{ $  0.303 \pm   0.067$}} \\ 
\hline\hline
\end{tabular}
\end{center}
\end{center}
\end{table}

Table~\ref{ffxvnim-dz-de-sub} provides the fractional dependence of the $\Bzero$
branching fractions listed in Table~\ref{xbrrsim-dz-de-smry} upon the $\Dzero$ branching fractions.
These allow corrections to the measurements to be made to account for improved determinations of the branching 
fractions~\cite{pdg} : ${\cal{B}}(\Dzero \to \Kpi) = ( 3.89 \pm 0.09) \%$, ${\cal{B}}(\Dzero \to
\Ktwopipl) = ( 13.9 \pm 0.9 )\% $ and ${\cal{B}}( \Dzero \to \Kthreepi ) = ( 7.49 \pm 0.31 )\% $. 
The coefficients are obtained from the subdecay branching fractions and the efficiencies for each subdecay. 
The fractional change to the branching fractions can be obtained from the product of the relative change of the subdecay
branching fraction and the corresponding negated coefficient.

\section{Systematic Uncertainties} 
\label{sec:systematics}
\mod{
Systematic uncertainties estimated for the $\Dzero\hzero$ and
$\Dstarzero\hzero$ modes for the combined $\Dzero$ submode samples  
are provided in Table~\ref{stotxim-dz-de-smry}. 
The systematic uncertainties are obtained by averaging those for the individual final
states using the PDG $\Dzero$ subdecay fractions. 
}
\mod{
The ``Tracking efficiency''
category accounts for charged products of $\Dstze$ only, estimated using an
uncertainty of 1\% per charged track.
The ``$\hzero$ efficiency'' category corresponds to 
efficiency uncertainties for $\pizero$, $\eta$ and $\omega$. These
uncertainties are obtained from the uncertainty on the efficiency
corrections which is estimated from comparisons of ratios of various processes
between data and Monte Carlo samples, as outlined in the previous section. 
The ``Extra \pizero'' category corresponds to the $\pizero$ from the $\Dzero \to \Ktwopi $ process.  
}

The cross-feed uncertainty is estimated as 25\% of the
contribution from this source in the signal regions.   
Uncertainties arising from the background and signal modelling used are estimated from the changes in the yields as a 
result of $\pm 1 \sigma$ variations on the model parameters.  

The ``Longitudinal polarization fraction'' uncertainty is relevant only for the vector-vector final state $\Dstarzero\omega$. 
Although the analysis does not directly use angular information in these
modes, the polarization can effect the orientation and hence momentum of the slow
$\pizero$ from the $\Dstarzero$ resulting in changes to the total efficiency.
Monte Carlo signal samples with various longitudinal polarization fractions ($f_L$) 
are used to estimate the size of the effect. The results are quoted assuming $f_L = 0.5 \pm 0.5 $; 
the variation in the efficiency over the full range of possible values is used
to estimate the systematic uncertainty.

\mod{
The dominant contributions arise from the background and signal modelling,
slow $\pizero$ efficiency for $\Dstarzero$ modes, and the subdecay branching
fractions.}
The total uncertainty is obtained  by summing the individual uncertainties in
quadrature.

\section{Conclusion}
Improved measurements of the branching fractions of the color-suppressed decays  $\Bzerobar \to \Dzero \pizero$, 
$ \Dzero \eta $, $ \Dzero \omega $, $\Dstarzero \pizero$, $\Dstarzero \eta$ and  $\Dstarzero \omega$ 
are presented. The results are consistent with the previous Belle measurements 
but with considerably improved precision due to the sevenfold increase in statistics.
\mod{
The individual results are consistent within two standard deviations with the BaBar
measurements~\cite{ref:Babar}; however it is notable that all the measurements
are lower than those of BaBar.
}
    
The measured values fall in the range (1.4-2.4) $\times 10^{-4}$,   
which is significantly higher than theoretical predictions based on naive factorization.
This discrepancy indicates that either final state rescattering is
significant, or else the assumption that the second Wilson coefficient
$a_{2}$ is real and process-independent is invalid~\cite{ref:Chua,ref:NeuPet}.

\section*{Acknowledgments}
We thank the KEKB group for the excellent operation of the
accelerator, the KEK cryogenics group for the efficient
operation of the solenoid, and the KEK computer group and
the National Institute of Informatics for valuable computing
and Super-SINET network support. We acknowledge support from
the Ministry of Education, Culture, Sports, Science, and
Technology of Japan and the Japan Society for the Promotion
of Science; the Australian Research Council and the
Australian Department of Education, Science and Training;
the National Science Foundation of China and the Knowledge Innovation Program
of Chinese Academy of Sciences under contract No.~10575109 and IHEP-U-503;
the Department of Science and Technology of
India; the BK21 program of the Ministry of Education of
Korea, and the CHEP SRC program and Basic Research program 
(grant No. R01-2005-000-10089-0) of the Korea Science and
Engineering Foundation; the Polish State Committee for
Scientific Research under contract No.~2P03B 01324; the
Ministry of Science and Technology of the Russian
Federation; the Slovenian Research Agency;  the Swiss National Science
Foundation; the National Science Council and
the Ministry of Education of Taiwan; and the U.S.\
Department of Energy.


\begin{thebibliography}{99}
\bibitem{CC}
Throughout this paper, 
the inclusion of the charge conjugate mode decay is implied
unless otherwise stated.
\end{thebibliography}
\end{document}